\documentclass{aa}

\def\kms{km\,s$^{-1}$}

\usepackage{graphicx}
\usepackage{longtable}
\usepackage{array}
\usepackage{amssymb}
\begin{document}

\title{Multiplicity among chemically peculiar stars}

\subtitle{II. Cool magnetic Ap stars\thanks{Based on observations collected at
the Observatoire de Haute-Provence (CNRS), France}$^,$\thanks{Tables 1 and 3
are only available in electronic form at the CDS via anonymous ftp to
cdsarc.u-strasbg.fr (130.79.128.5) or via
http://cdsweb.u-strasbg.fr/Abstract.html}}

\author{F. Carrier\inst{1}
        \and P. North\inst{2}
	\and S. Udry\inst{1}
        \and J. Babel\inst{3}}
   
\institute{Observatoire de Gen\`eve, CH-1290 Sauverny, Switzerland
\and Institut d'Astronomie de l'Universit\'e de Lausanne, 
 CH-1290 Chavannes-des-bois, Switzerland
\and Office F\'ed\'eral de la Statistique, Espace de l'Europe 10,
CH-2010 Neuch\^atel, Switzerland}

\offprints{P.~North}

\date{Received  / Accepted }

\titlerunning{Multiplicity among CP stars}
\authorrunning{Carrier et al.}

\abstract{
We present new orbits for sixteen Ap spectroscopic binaries, four of which
might in fact be Am stars, and give their orbital elements. Four of them are 
SB2 systems: HD~5550, HD~22128, HD~56495 and HD~98088.
The twelve other stars are : HD~9996, HD~12288,
HD~40711, HD~54908, HD~65339, HD~73709, HD~105680, HD~138426,
HD~184471, HD~188854, HD~200405 and HD~216533. Rough estimates of the individual
masses of the components
of HD~65339 ($53$ Cam) are given, combining
our radial velocities with the results of speckle interferometry and with
Hipparcos parallaxes.
Considering the mass functions of 74 spectroscopic binaries from this work
and from the literature,
we conclude that the distribution of the mass ratio is the same for cool
Ap stars as for normal G dwarfs. Therefore, the only differences between
binaries with normal stars and those hosting an Ap star lie in the period 
distribution: except for the case of HD 200405, all orbital periods are
longer than (or equal to) 3 days. A consequence of this peculiar distribution
is a deficit of null eccentricities. There is no indication that the secondary
has a special nature, like e.g. a white dwarf.
}

\maketitle

\section{Introduction}
Ap stars are conspicuous not only because of their strong chemical anomalies,
but also because of their strong, large-scale magnetic field (at least in
the Si and SrCrEu subtypes) and slow rotation. The latter characteristic
is associated with a complete lack of Ap stars in binaries with very short
orbital periods (i.e. 1.5 days or less), contrarily to normal stars,
probably because such systems are
synchronized and their components have to rotate fast, which does not seem
compatible with the development of chemical peculiarities.
One could also think of an observational bias as another possible cause,
the line widening erasing mild peculiarities; the fact that some Bp or Ap
stars do rotate fast (200~km\,s$^{-1}$) does not support this explanation,
however. But, in addition
to the fact that tidal synchronisation will preclude  the existence of Ap stars
in short period systems, one may reasonably expect that some special conditions
are needed to form an Ap star, and that these conditions might leave their
blueprint not only in the magnetic field and slow rotation, but also in the
frequency and orbital elements of binaries. One important purpose of this
paper is precisely to explore this possibility.

The first, systematic search for binaries among Ap stars has been done by
Abt \& Snowden (\cite{AS73}), who examined 62 bright northern stars and concluded
to a low rate of binaries (20 percent), except for HgMn stars (43 percent).
Aikman (\cite{A76}) increased the sample of HgMn stars from 15 to 80 and confirmed
the rate found by Abt \& Snowden, since he found 49 percent, which is very
close to the result of Jaschek \& Gomez (\cite{JG70}) for normal B0 to M stars
of the main sequence ($47\pm 5 \%
$).

There has been no systematic review of multiplicity among Ap stars since
the work of Gerbaldi et al. (\cite{GFH85}, hereafter GFH85), apart from some attempts 
by Budaj (\cite{B95}, \cite{B96}, \cite{B97}) to interpret the role that a binary companion
might have in the appearance of chemical peculiarities of both Am and Ap stars.
According to GFH85, the rate of binaries tends to
be rather small among the He-weak and Si stars. For the coolest Ap stars, as 
well as for the HgMn stars, this rate behaves in the same way as for normal 
stars. Moreover, the magnetic Ap stars show a strong deficit of SB2 binaries:
Only two SB2's containing a magnetic Ap star had been well studied before the
CORAVEL observations: HD~55719 (Bonsack \cite{B76}) and HD~98088
(Abt et al. \cite{A68}; Wolff \cite{W74}). On the other hand, HgMn stars, which generally
have no significant magnetic field (see, however, Mathys \& Hubrig \cite{MH95}), are
often found in SB2 systems, and their companion seems to be always an Am star
when its effective temperature is below 10000~K (Ryabchikova \cite{Ry98}).
Am stars are also known to be frequently associated with SB2 systems
(Abt \& Levy \cite{A85}).

A radial-velocity survey of a small number of cool, well-known magnetic Ap
stars has been initiated in 1980 using the CORAVEL scanner (Baranne et al.
\cite{BMP79}), and the sample
has been extended in 1985 to all stars brighter than $V=8.6$, visible from
the northern hemisphere and having Geneva photometry. The purpose was to
increase the relatively poor statistics and obtain a better understanding
of the role multiplicity might play in the context of chemically peculiar stars.
Preliminary results have been published by North (\cite{N94}), especially the 
discovery of a long-period SB2 system which has been studied in more details 
later (HD 59435, Wade et al. \cite{WN96}, \cite{WM99}) and the discovery of an
SB1 system with a period as short as 1.6 days (HD 200405). 

This paper is the second one in a series dedicated to multiplicity among
Ap and Am stars. Since Drs. Nicole Ginestet and
Jean-Marie Carquillat in Toulouse had independantly measured a few of our
programme stars with the same instrument, we had dedicated the first paper
of this series to four common stars (North et al. \cite{NC98} Paper I),
especially the Ap
stars HD~8441 and $\beta$ CrB. Here we present the results for all Ap stars 
measured to date with CORAVEL, with a few additional data from the ELODIE
spectrograph (Baranne et al. \cite{BQ96}) which are by-products of a survey of
magnetic fields (Babel \& North, in preparation).

\section{Observations and sample}

All radial-velocity observations have been obtained at the Observatoire de
Haute-Provence. Most of the data were measured with the CORAVEL scanner attached
to the 1-meter Swiss telescope. Although this instrument is optimized for
late-type stars, it can still yield very good results on slowly rotating F and
even A stars, especially if their metallic lines are enhanced, as is the case
of Ap stars. Originally, the sample was selected according to the following
criteria:
$T_{\mbox{eff}} \leqslant 10000$~K according to Geneva photometry
(stars for which no Geneva photometry existed in 1985 were not retained),
$\delta \geqslant -25\degr$, $m_V \leqslant 8.6$ (except for a few
stars lying in the northern galactic polar cap and for candidate
high-velocity stars proposed by Jaschek et al. \cite{JJ83}). 
A total of 1913 radial-velocity observations were made of the 119 programme 
stars during the period March 1980 through April 1998. Errors for the individual
observations are derived following the precepts of Baranne et al. (\cite{BMP79})
and are generally below 2.0 \kms\, for all stars.
The radial-velocity variations are large enough
(10 \kms $< 2\times K < 173$ \kms) to be clearly significant.
In addition to the CORAVEL data, we use here 75 measurements for 72 stars
already observed with CORAVEL, obtained with the 
1.93-meter telescope at OHP, equipped with the ELODIE spectrograph.
This fibre-fed echelle spectrograph has been in operation
only since the end of 1993 and has a better precision than the CORAVEL scanner.
All but one ELODIE data were gathered during four observing runs from October 1994 to
September 1996, which were primarily aimed at determining magnetic fields
(Babel et al. \cite{BNQ95}, Babel \& North \cite{BN97}); the radial velocities are just a
by-product of the programme. One additional data was obtained in April 2002
for HD 116114 with ELODIE.

Among all the programme stars, only 20 stars have as yet a precise orbit.
$\beta$~CrB and HD 8441 (as well as the Am star HD 43478) have been published by
North et al. (\cite{NC98}), while HD 59435 and HD 81009 have been published by
Wade et al. (\cite{WN96}, \cite{WM99}) and by Wade et al. (\cite{WD00})
respectively; the orbit of HD 73709 was published by Debernardi et al.
(\cite{DM20}). The others are presented in the appendix A, as well as an
improved orbit of HD 73709. The orbital elements of the 16
stars studied in this article are listed in Table~\ref{orbel}.
The individual $RV$ measurements of all stars of
the sample\footnote{except for HD 43478, HD 59435, HD 81009 and $\beta$~CrB
which were already published. The data for HD 8441 are given anew in Tables 1
and 2, because a few additional measurements have been made and all measurements
are now in the ELODIE $RV$ system; these data may be useful since HD 8441 is a
triple system whose longer period is not yet known.}, as well as the depths and
widths of the correlation dips, are listed
in Table~1, which is only available in electronic form. The average radial
velocities, upper limits to projected rotational velocities and some statistical
quantities like $P(\chi^2)$ (see Duquennoy \& Mayor \cite{DM91} for a definition
of $P(\chi^2)$), are given in Table \ref{tab2}. The list of the 48 stars
observed with CORAVEL (in addition to the 119 stars mentioned above), but
showing no measurable correlation dip is given in
Table~3 (only available in electronic form).

\setcounter{table}{1}
\onecolumn
\setlength{\tabcolsep}{5pt}
\setlongtables
\begin{small}
\begin{longtable}{|l|l|l|r|r|r|r|r|r|r|r|r|r|l|} 
\caption{CORAVEL's sample with ST: spectral type, Johnson B-V index, RV: mean radial velocity,
SRVM: standard deviation on the mean RV, SCAT: external scatter (sigma) of the RV values, E/I:
ratio of external to internal errors, N: number of CORAVEL measurements used
in the statistics, SPAN: time span of the CORAVEL measurements, upper limit to $v \sin
i$, SVS: 
standard deviation on $v \sin i$ (undefined if 99.9), P($\chi^2$): probability of no intrinsic RV
variation (undefined if 9.999) and remark. (1) North et al. \cite{NC98}; (2) Mkrtichian et al.
\cite{MS98}; (3) Northcott \cite{N48}.}
\\
\hline
\multicolumn{1}{|c|}{\label{tab2}HD}& \multicolumn{1}{c|}{Renson} & \multicolumn{1}{c|}{ST}  & \multicolumn{1}{c|}{B-V} &
 \multicolumn{1}{c|}{RV} & \multicolumn{1}{c|}{SVRM} &
 \multicolumn{1}{c|}{SCAT} & \multicolumn{1}{c|}{E/I} &
 \multicolumn{1}{c|}{N} & \multicolumn{1}{c|}{SPAN} & \multicolumn{1}{c|}{$v \sin i$} &
 \multicolumn{1}{c|}{SVS} & \multicolumn{1}{c|}{P($\chi^2$)} &
 \multicolumn{1}{c|}{Rem}  \\ 
\hline
\endfirsthead
\hline
\multicolumn{1}{|c|}{HD}& \multicolumn{1}{c|}{Renson} & \multicolumn{1}{c|}{ST}  & \multicolumn{1}{c|}{B-V} &
 \multicolumn{1}{c|}{RV} & \multicolumn{1}{c|}{SVRM} &
  \multicolumn{1}{c|}{SCAT} & \multicolumn{1}{c|}{E/I} &
 \multicolumn{1}{c|}{N} & \multicolumn{1}{c|}{SPAN} & \multicolumn{1}{c|}{$v \sin i$} &
 \multicolumn{1}{c|}{SVS} & \multicolumn{1}{c|}{P($\chi^2$)} &
 \multicolumn{1}{c|}{Rem}  \\
\hline
\endhead
\hline
\endfoot
\hline
2453	&560	& A1 Sr Eu Cr	  &  0.08    &       -18.16  &       0.17    &       0.71    &       1.41    &       18      &       5578    &       4.4     &       0.5     &       0.009   &       SB1, spot?  \\
2957	&760	& B9 Cr Eu	  &  0.02    &       12.32   &       0.85    &       1.35    &       0.65    &       6       &       4024    &       24.4    &       2.4     &       0.853   &        \\
5550	&1470	& A0 Sr 	  &  -0.02   &       -5.80   &       5.00    &       1.79    &       13.77   &       19      &       385     &       6.5     &       1.3     &       0.000   &       orbit  \\
5797	&1530	& A0 Cr Eu Sr	  &  0.26    &       -6.32   &       0.16    &       0.54    &       1.14    &       12      &       5873    &       3.0     &       0.9     &       0.224   &        \\
8441	&2050	& A2 Sr 	  &  0.01    &       7.65    &       1.64    &       7.32    &       18.73   &       112     &       6645    &       2.9     &       0.6     &       0.000   &       orbit (1) \\
9996	&2470	& B9 Cr Eu Si	  &  -0.02   &       3.98    &       1.31    &       8.47    &       7.23    &       42      &       3677    &       $<$2.0  &       1.7     &       0.000   &       orbit  \\
10088	&2520	& A1-F1 	  &  0.31    &       11.87   &       0.68    &       2.45    &       1.20    &       13      &       4791    &       65.8    &       6.6     &       0.160   &        \\
11187	&2770	& A0 Si Cr Sr	  &  -0.07   &       7.77    &       1.30    &       1.23    &       0.42    &       5       &       4016    &       16.9    &       1.2     &       0.950   &        \\
12288	&3130	& A2 Cr Si	  &  0.08    &       -52.51  &       1.18    &       6.47    &       7.29    &       30      &       5948    &       12.5    &       0.4     &       0.000   &       orbit  \\
15089	&3760	& A4 Sr 	  &  0.12    &       3.66    &       2.65    &       5.30    &       1.38    &       4       &       2733    &       36.0    &       6.8     &       0.137   &        \\
15144	&3770	& A5 Sr Cr Eu	  &  0.15    &       -0.56   &       0.62    &       0.62    &       1.00    &       1       &       0       &       11.2    &       0.7     &       9.999   &        \\
16145	&4060	& A0 Cr Sr Eu	  &  0.05    &       7.57    &       1.41    &       2.82    &       1.40    &       4       &       1021    &       27.9    &       2.9     &       0.137   &        \\
17775	&4430	& A1 Cr Eu	  &  0.14    &       -6.00   &       0.23    &       0.79    &       0.89    &       15      &       3533    &       14.9    &       0.4     &       0.702   &        \\
18078	&4500	& A0 Sr Cr	  &  0.21    &       -16.97  &       0.59    &       2.99    &       5.28    &       26      &       6426    &       4.0     &       0.5     &       0.000   &       SB1?  \\
18296	&4560	& A0 Si Sr	  &  -0.01   &       2.42    &       4.20    &       4.20    &       1.00    &       1       &       0       &       28.5    &       3.7     &       9.999   &        \\
22128 a	&5560	& A7 Sr Eu Mn	  &  0.34    &       23.16   &       3.09    &       3.98    &       75.27   &       17      &       632     &       15.9    &       0.3     &       0.000   &       orbit  \\
22128 b	&	&		  &  0.34    &        8.93	  &	  5.04    &	  0.16    &	  42.95   &	  16	  &	  632	  &	  16.3    &	  0.7	  &	  0.000   &	   \\
22374	&5660	& A1 Cr Sr Si	  &  0.12    &       0.08    &       0.29    &       0.88    &       1.39    &       9       &       4357    &       5.2     &       0.9     &       0.058   &        \\
23207	&5900	& A2 Sr Eu	  &  0.19    &       -0.13   &       0.24    &       0.33    &       0.55    &       6       &       1402    &       6.3     &       1.1     &       0.916   &        \\
24712	&6320	& A9 Sr Eu Cr	  &  0.33    &       23.14   &       0.12    &       1.44    &       1.16    &       156     &       1575    &       6.5     &       0.4     &       0.011   &       roAp \\
25163	&6400	& A2 Sr Cr	  &  0.15    &       26.37   &       1.23    &       4.25    &       2.61    &       12      &       2210    &       21.7    &       2.2     &       0.000   &       spot?  \\
25163 a	&	&		  &  0.15 &	  37.66   &	  1.25    &	  1.25    &	  1.00    &	  1	  &	  0	  &	  13.5    &	  6.4	  &	  9.999   &	   \\
25163 b	&	&		  &  0.15 &	  12.62   &	  2.24    &	  2.24    &	  1.00    &	  1	  &	  0	  &	  14.6    &	  9.4	  &	  9.999   &	  \\
25354	&6460	& A2 Eu Cr	  &  0.04    &       -11.33  &       1.16    &       2.83    &       1.40    &       6       &       3902    &       15.0    &       1.9     &       0.091   &        \\
35353	&9030	& A0 Sr Cr Eu	  &  0.22    &       18.28   &       0.58    &       1.18    &       0.83    &       6       &       3616    &       19.4    &       1.0     &       0.639   &        \\
38104	&10240  & A1 Cr Eu	  &  0.03    &       -5.01   &       1.42    &       4.02    &       1.66    &       8       &       3211    &       25.2    &       2.5     &       0.008   &       spot?  \\
38823	&10440  & A5 Sr Eu	  &  0.39    &       -6.85   &       0.32    &       0.50    &       0.64    &       6       &       3384    &       12.0    &       0.9     &       0.857   &        \\
40711	&10880  & A0 Sr Cr Eu	  &  0.15    &       -7.63   &       1.10    &       5.59    &       12.05   &       26      &       2570    &       $<$2.0  &       0.9     &       0.000   &       orbit  \\
41403	&11080  & B9 Sr Cr Eu	  &  0.02    &       0.74    &       0.46    &       1.83    &       1.27    &       16      &       2469    &       21.6    &       2.2     &       0.074   &        \\
42326	&11280  & A0 Eu Cr	  &  0.06    &       25.82   &       0.60    &       1.48    &       1.35    &       6       &       2509    &       14.6    &       0.9     &       0.105   &        \\
42616	&11390  & A1 Sr Cr Eu	  &  0.09    &       1.53    &       1.18    &       3.54    &       1.86    &       9       &       4465    &       18.7    &       0.9     &       0.001   &       spot?  \\
42777	&11450  & A0 Eu Sr Cr	  &  0.03    &       22.56   &       0.99    &       1.98    &       1.33    &       4       &       2135    &       18.5    &       2.0     &       0.164   &        \\
47103	&12630  & A  Sr Eu	  &  -0.01   &       -14.37  &       0.99    &       0.99    &       1.00    &       1       &       0       &       24.3    &       2.4     &       9.999   &        \\
49976	&13560  & A1 Sr Cr Eu	  &  0.01    &       19.39   &       0.90    &       4.02    &       1.58    &       20      &       3741    &       34.2    &       3.4     &       0.000   &       spot?  \\
50169	&13700  & A3 Sr Cr Eu	  &  -0.05   &       13.17   &       0.48    &       1.45    &       1.99    &       9       &       2541    &       7.0     &       1.1     &       0.000   &       SB1?  \\
52696	&14460  & A3 Sr Eu Cr	  &  0.15    &       32.29   &       0.79    &       2.37    &       3.43    &       9       &       2262    &       3.4     &       2.0     &       0.000   &       SB1  \\
54908	&15000  & A0 Si 	  &  0.29    &       5.55    &       3.76    &       7.22    &       8.72    &       21      &       2922    &       55.2    &       5.5     &       0.000   &       orbit  \\
56495	&15430  & A3p		  &  0.34    &       -8.72   &       0.94    &       0.35    &       0.26    &       2       &       4396    &       30.8    &       8.0     &       0.795   &       orbit  \\
56495 a	&	&		  &  0.34 &	  -7.60   &	  4.44    &	  1.28    &	  13.92   &	  23	  &	  4396    &	  25.3    &	  2.5	  &	  0.000   &	   \\
56495 b	&	&		  &  0.34 &	  -2.08   &	  6.38    &	  9.24    &	  18.25   &	  21	  &	  2507    &	  12.5    &	  2.1	  &	  0.000   &	   \\
62140	&17050  & A8 Sr Eu	  &  0.26    &       7.95    &       0.40    &       0.68    &       0.56    &       9       &       4701    &       21.1    &       2.1     &       0.963   &       SB1? (2) \\
65339	&17910  & A3 Sr Eu Cr	  &  0.14    &       -0.27   &       1.11    &       6.86    &       7.26    &       38      &       5084    &       19.4    &       0.6     &       0.000   &       orbit  \\
71866	&19850  & A1 Eu Sr Si	  &  0.09    &       28.33   &       0.41    &       1.24    &       1.60    &       9       &       4464    &       14.9    &       0.8     &       0.010   &       spot?  \\
72968	&20240  & A2 Sr Cr	  &  -0.03   &       24.81   &       0.53    &       1.26    &       0.97    &       6       &       2329    &       15.9    &       1.4     &       0.473   &        \\
73709	&20510  & A3-F2 	  &  -0.01   &       41.73   &       3.06    &       8.47    &       26.98   &       2       &       4       &       16.0    &       0.9     &       0.000   &       orbit  \\
74521	&20790  & A1 Si Eu Cr	  &  -0.10   &       25.54   &       0.53    &       2.12    &       1.03    &       16      &       5143    &       16.8    &       0.7     &       0.397   &        \\
77350	&21860  & B9 Sr Cr Hg	  &  -0.04   &       -19.01  &       1.55    &       4.40    &       1.44    &       8       &       5876    &       19.5    &       1.6     &       0.058   &        \\
82093	&23340  & A2 Sr Eu Cr	  &  0.11    &       -10.27  &       0.77    &       2.65    &       1.52    &       12      &       3739    &       21.4    &       2.1     &       0.012   &       spot?  \\
86170	&24620  & A2 Sr Cr Eu	  &  0.13    &       13.39   &       0.76    &       2.85    &       5.13    &       14      &       2570    &       $<$2.0  &       1.4     &       0.000   &       spot,SB1?  \\
89069	&25500  & A0 Sr Cr Eu	  &  0.63    &       -9.78   &       0.25    &       0.42    &       0.59    &       8       &       2539    &       4.3     &       1.1     &       0.934   &        \\
90569	&26010  & A0 Sr Cr Si	  &  -0.06   &       -7.81   &       1.57    &       1.57    &       1.00    &       1       &       0       &       7.0     &       3.1     &       9.999   &        \\
94427	&27250  & A5 Sr 	  &  0.31    &       17.71   &       0.20    &       0.32    &       0.66    &       6       &       2325    &       $<$2.0  &       0.0     &       0.825   &        \\
98088	&28310  & A8 Sr Cr Eu	  &  0.21    &       0.71    &       6.30    &       6.11    &       42.04   &       8       &       2340    &       22.8    &       2.3     &       0.000   &       orbit  \\
98088 a	&	&		  &  0.21 &	  7.59    &	  6.96    &	  8.75    &	  54.85   &	  12	  &	  1531    &	  21.1    &	  2.1	  &	  0.000   &	   \\
98088 b	&	&		  &  0.21 &	  -36.84  &	  0.21    &	  0.01    &	  33.36   &	  12	  &	  1531    &	  15.8    &	  1.9	  &	  0.000   &	   \\
99563	&28660  & F0 Sr 	  &  0.28    &       -1.01   &       2.55    &       2.55    &       1.00    &       1       &       0       &       30.1    &       3.9     &       9.999   &        \\
102333	&29500  & A0 Eu Cr Sr	  &  0.15    &       -5.06   &       0.47    &       0.47    &       1.00    &       1       &       0       &       $<$2.0  &       0.0     &       9.999   &        \\
105680	&30570  & A2-F1 	  &  0.30    &       -9.63   &       3.49    &       2.64    &       36.06   &       42      &       2966    &       14.1    &       0.2     &       0.000   &       orbit  \\
107612	&31190  & A2 Sr 	  &  0.05    &       14.13   &       3.16    &       6.32    &       1.58    &       4       &       1914    &       37.9    &       3.8     &       0.076   &        \\
110066	&31960  & A1 Sr Cr Eu	  &  0.06    &       -12.94  &       0.21    &       0.19    &       0.38    &       6       &       2542    &       5.2     &       0.6     &       0.983   &        \\
111133	&32310  & A1 Sr Cr Eu	  &  -0.05   &       18.82   &       0.32    &       0.58    &       0.75    &       6       &       2543    &       7.5     &       0.5     &       0.732   &        \\
112528	&32730  & A3 Sr Eu Cr	  &  0.32    &       -14.43  &       0.39    &       1.11    &       1.03    &       8       &       4452    &       20.9    &       2.1     &       0.398   &        \\
113894	&33000  & A7 Sr Cr Eu	  &  -0.02   &       7.94    &       0.81    &       0.81    &       1.00    &       1       &       0       &       15.7    &       1.8     &       9.999   &        \\
115708	&33450  & A3 Sr Eu	  &  0.25    &       2.90    &       0.33    &       0.81    &       0.87    &       8       &       4343    &       11.0    &       0.7     &       0.637   &        \\
116114	&33530  & F0 Sr Cr Eu	  &  0.30    &       4.67    &       0.53    &       2.42    &       4.74    &       21      &       4414    &       8.5     &       0.3     &       0.000   &       SB1  \\
118022	&34020  & A2 Cr Eu Sr	  &  0.03    &       -7.45   &       0.28    &       1.26    &       1.77    &       20      &       4385    &       10.0    &       0.4     &       0.000   &       spot  \\
119213	&34410  & A3 Sr Cr	  &  0.11    &       -5.15   &       1.37    &       3.37    &       1.16    &       6       &       2126    &       36.1    &       3.6     &       0.259   &        \\
122208	&35050  & A2 Sr Cr Eu	  &  0.13    &       5.66    &       1.10    &       1.10    &       1.00    &       1       &       0       &       30.3    &       3.8     &       9.999   &        \\
122569	&35180  & A0 Cr Eu Sr	  &  0.05    &       5.87    &       1.01    &       1.01    &       1.00    &       1       &       0       &       19.0    &       2.9     &       9.999   &        \\
124437	&35620  & A0 Cr Sr Eu	  &  0.63    &       -21.49  &       2.39    &       2.39    &       1.00    &       1       &       0       &       31.2    &       5.1     &       9.999   &        \\
125248	&35760  & A1 Eu Cr	  &  -0.01   &       -7.88   &       0.75    &       0.52    &       0.49    &       2       &       2       &       11.5    &       3.2     &       0.625   &        \\
126515	&36050  & A2 Cr Sr	  &  0.01    &       -2.92   &       0.22    &       1.05    &       0.90    &       28      &       5881    &       19.1    &       0.9     &       0.757   &        \\
127241	&36210  & A0 Sr Cr	  &  -0.09   &       0.21    &       1.13    &       1.13    &       1.00    &       1       &       0       &       8.1     &       4.7     &       9.999   &        \\
128898	&36710  & A9 Sr Eu	  &  0.24    &       6.84    &       0.15    &       0.80    &       1.14    &       30      &       3       &       13.5    &       0.3     &       0.130   &        \\
130559	&37160  & A1 Sr Cr Eu	  &  0.07    &       -2.89   &       0.69    &       1.44    &       0.78    &       7       &       4312    &       21.4    &       2.1     &       0.731   &        \\
133029	&37770  & B9 Si Cr Sr	  &  -0.14   &       -10.01  &       1.06    &       2.69    &       0.90    &       8       &       4385    &       19.3    &       1.7     &       0.627   &        \\
134214	&38100  & F2 Sr Eu Cr	  &  0.35    &       -14.40  &       0.19    &       0.35    &       0.64    &       8       &       4339    &       2.7     &       1.2     &       0.899   &        \\
134793	&38240  & A4 Sr Eu Cr	  &  0.14    &       -1.48   &       0.70    &       1.84    &       1.12    &       7       &       2855    &       26.4    &       2.6     &       0.279   &        \\
135297	&38390  & A0 Sr Cr Eu	  &  -0.01   &       -35.72  &       0.76    &       2.10    &       0.97    &       8       &       3620    &       16.9    &       1.4     &       0.493   &        \\
137949	&39240  & F0 Sr Eu Cr	  &  0.38    &       -28.61  &       0.25    &       0.91    &       1.72    &       13      &       5881    &       7.5     &       0.5     &       0.000   &       spot  \\
138426	&39420  & B9 Sr Cr	  &  0.12    &       -12.86  &       7.73    &       5.65    &       33.11   &       11      &       2290    &       $<$2.0  &       1.8     &       0.000   &       orbit  \\
140160	&39840  & A1 Sr 	  &  0.03    &       16.52   &       3.15    &       3.15    &       1.00    &       1       &       0       &       14.1    &       5.8     &       9.999   &        \\
142070	&40330  & A0 Sr Cr Eu	  &  0.15    &       -8.58   &       0.31    &       1.04    &       1.45    &       11      &       2289    &       8.1     &       0.8     &       0.022   &       SB1  \\
146971	&41510  & A0 Sr Cr Eu	  &  0.22    &       -2.47   &       1.21    &       2.97    &       1.10    &       6       &       2855    &       25.8    &       2.6     &       0.307   &        \\
148330	&41910  & A2 Si Sr	  &  0.63    &       -3.55   &       0.79    &       1.93    &       1.55    &       6       &       2218    &       8.2     &       2.1     &       0.041   &        \\
149911	&42420  & A0 Cr Si Sr	  &  0.16    &       -18.70  &       0.68    &       1.62    &       0.85    &       8       &       3728    &       23.2    &       2.3     &       0.658   &        \\
152107	&43050  & A3 Sr Cr Eu	  &  0.09    &       -0.42   &       0.58    &       0.82    &       0.64    &       5       &       2218    &       19.5    &       0.4     &       0.804   &        \\
153882	&43450  & A1 Cr Eu	  &  0.04    &       -28.69  &       0.62    &       1.43    &       0.94    &       6       &       2217    &       19.9    &       1.2     &       0.497   &        \\
157740	&44300  & A3 Cr Eu Sr	  &  0.07    &       12.22   &       6.39    &       2.79    &       3.34    &       4       &       1805    &       42.5    &       6.4     &       0.000   &       spot?  \\
165474	&46650  & A7 Sr Cr Eu	  &  0.31    &       13.45   &       0.24    &       0.40    &       0.67    &       6       &       1531    &       9.2     &       0.7     &       0.813   &        \\
168481	&47220  & A7 Sr Cr	  &  0.28    &       -3.90   &       0.64    &       1.12    &       0.62    &       8       &       4028    &       34.1    &       3.4     &       0.917   &        \\
168796	&47310  & A0 Si Cr Sr	  &  0.11    &       12.29   &       0.51    &       2.82    &       3.36    &       31      &       4707    &       14.3    &       0.3     &       0.000   &       SB1  \\
171586	&48090  & A2 Sr Cr	  &  0.07    &       -0.62   &       2.92    &       7.74    &       2.30    &       7       &       2042    &       31.9    &       3.5     &       0.000   &       spot?  \\
172032	&48210  & A9 Sr Cr	  &  0.33    &       -14.37  &       0.19    &       0.45    &       0.89    &       7       &       3786    &       3.6     &       1.0     &       0.589   &        \\
173650	&48660  & A0 Si Sr Cr	  &  0.02    &       -14.59  &       0.89    &       0.73    &       0.37    &       5       &       1983    &       10.3    &       2.1     &       0.971   &        \\
176232	&49160  & A6 Sr 	  &  0.25    &       18.14   &       0.15    &       0.27    &       0.59    &       9       &       2993    &       2.2     &       0.8     &       0.950   &        \\
180058	&50000  & A3 Sr 	  &  0.34    &       -2.80   &       1.82    &       4.81    &       1.96    &       7       &       3787    &       27.3    &       2.7     &       0.001   &       spot?  \\
180778	&50150  & A2p		  &  0.15    &       -24.20  &       0.33    &       0.80    &       1.25    &       6       &       2898    &       12.7    &       0.5     &       0.167   &        \\
184471	&50890  & A9 Sr Cr Eu	  &  0.29    &       -26.38  &       1.98    &       1.55    &       19.36   &       34      &       3418    &       $<$2.0  &       0.0     &       0.000   &       orbit  \\
186343	&51370  & A3-F0 	  &  0.23    &       -19.05  &       0.17    &       0.42    &       0.85    &       8       &       4425    &       8.9     &       0.5     &       0.658   &        \\
188041	&51900  & A6 Sr Cr Eu	  &  0.19    &       -21.74  &       0.19    &       0.35    &       0.70    &       7       &       1981    &       2.9     &       0.7     &       0.823   &        \\
188593	&52110  & A5-F5 Sr	  &  0.30    &       -25.64  &       0.17    &       0.50    &       0.98    &       9       &       4420    &       8.3     &       0.6     &       0.485   &        \\
188854	&52220  & A7p:  	  &  0.31    &       -25.89  &       5.33    &       0.63    &       68.74   &       33      &       4424    &       9.3     &       0.2     &       0.000   &       orbit  \\
190145	&52790  & A2p		  &  0.25    &       -14.48  &       0.29    &       0.90    &       1.65    &       10      &       4424    &       11.7    &       0.3     &       0.004   &        \\
191654	&53420  & A2 Sr Cr	  &  0.21    &       -15.78  &       0.36    &       1.86    &       1.51    &       27      &       4794    &       23.4    &       2.3     &       0.000   &       SB1,spot?  \\
191742	&53440  & A5 Sr Cr	  &  0.22    &       -3.11   &       0.15    &       0.56    &       1.17    &       14      &       5913    &       $<$2.0  &       0.0     &       0.177   &        \\
192224	&53580  & A2 Cr Eu	  &  0.07    &       -23.44  &       0.79    &       3.36    &       2.37    &       18      &       4412    &       21.7    &       2.2     &       0.000   &       spots  \\
192678	&53740  & A2 Cr 	  &  0.03    &       -38.46  &       0.18    &       0.24    &       0.48    &       8       &       2752    &       7.0     &       0.4     &       0.978   &        \\
192913	&53840  & A0 Si Cr	  &  -0.07   &       -8.77   &       1.11    &       2.94    &       1.21    &       7       &       3230    &       14.3    &       1.7     &       0.193   &        \\
196133	&54660  & A1 Si Sr	  &  0.04    &       -34.11  &       1.20    &       5.03    &       7.36    &       5       &       684     &       22.6    &       4.7     &       0.000   &       orbit (3) \\
196502	&54780  & A2 Sr Cr Eu	  &  0.07    &       10.98   &       0.32    &       1.37    &       2.40    &       18      &       6590    &       8.3     &       0.3     &       0.000   &       SB1?  \\
199180	&55460  & A0 Si Cr	  &  0.03    &       -16.29  &       0.36    &       0.60    &       0.63    &       7       &       2501    &       2.3     &       2.6     &       0.887   &        \\
200405	&55830  & A2 Sr Cr	  &  0.09    &       0.18    &       0.94    &       5.41    &       7.00    &       33      &       3783    &       9.6     &       0.4     &       0.000   &       orbit  \\
201174	&56130  & A1 Cr Eu Sr	  &  0.02    &       -11.37  &       0.40    &       0.98    &       1.10    &       6       &       3332    &       17.4    &       1.1     &       0.316   &        \\
201601	&56210  & A9 Sr Eu	  &  0.26    &       -16.46  &       0.15    &       0.52    &       1.11    &       13      &       5917    &       4.6     &       0.4     &       0.250   &        \\
204411	&56920  & A6 Cr 	  &  0.08    &       -13.69  &       0.15    &       0.58    &       0.99    &       15      &       5917    &       5.3     &       0.6     &       0.481   &        \\
206088	&57390  & A8-F4 Sr	  &  0.32    &       -35.89  &       1.12    &       5.47    &       3.50    &       24      &       6654    &       45.3    &       4.5     &       0.000   &       SB, spot?  \\
213232	&59100  & A4 Sr 	  &  0.14    &       -23.34  &       0.51    &       2.21    &       1.14    &       19      &       4026    &       23.5    &       2.3     &       0.196   &        \\
216533	&59810  & A1 Sr Cr	  &  0.08    &       -5.04   &       0.51    &       3.48    &       5.44    &       47      &       6225    &       5.7     &       0.3     &       0.000   &       orbit  \\
216931	&59880  & A0		  &  0.21    &       6.99    &       7.98    &       1.29    &       2.87    &       2       &       1456    &       32.9    &       99.9    &       0.004   &       spot?  \\
220825	&60520  & A1 Cr Sr Eu	  &  0.04    &       -11.75  &       4.61    &       0.30    &       2.72    &       5       &       680     &       39.2    &       6.0     &       0.000   &       spot?  \\
221394	&60670  & A0 Sr Cr Si	  &  0.03    &       -8.83   &       3.35    &       7.50    &       1.47    &       5       &       2253    &       65.7    &       6.6     &       0.092   &        \\
221568	&60730  & A1 Sr Cr Eu	  &  0.11    &       -7.54   &       0.24    &       0.86    &       0.96    &       14      &       5967    &       $<$2.0  &       1.9     &       0.561   &        \\
\end{longtable}

\end{small}
\twocolumn
\section{The correlation dip of magnetic Ap stars}

It is interesting to discuss the properties of the correlation dip in the
case of magnetic Ap stars. For normal stars, this dip yields essentially three
independant informations, corresponding to the three parameters of the fitted
gaussian: the radial velocity, the width ($\sigma$) of the dip which is linked
with the projected rotational velocity $v\sin i$, and ``surface'' or
equivalent width $W$ of the dip, which depends on the effective temperature
and metallicity of the star (at least on the main sequence). A calibration
of the width $\sigma$ in terms of $v\sin i$ was
proposed by Benz \& Mayor (\cite{BM81}, \cite{BM84})
who showed that a temperature (or colour index $B-V$) term has to be introduced.
This calibration has been verified in the range of spectral types F6 to M0
(for luminosity class V),
but has been also applied to our hotter stars, implying a slight extrapolation.
Mayor (\cite{M80}) first showed that the dip's equivalent width $W$ could be
efficiently calibrated in terms of metallicity [Fe/H], provided a good
indicator of $T_{\mbox{eff}}$ be known, e.g. a colour index like Johnson's $B-V$
or Geneva $B2-V1$. An explicit calibration of $W$ for Am to G stars in terms
of [Fe/H] was first proposed by North \& Duquennoy (\cite{ND91}) and yielded good
results (the rms scatter of the difference between the spectroscopic and
CORAVEL [Fe/H] amounted to no more than 0.12 dex), although the sources of
spectroscopic [Fe/H] values were heterogeneous. Later, Pont (\cite{P97}) based a
similar calibration on the very large and homogeneous sample of spectroscopic
metallicities of Edvardsson et al. (\cite{E93}) and obtained an excellent result,
the rms scatter of the differences between spectroscopic and CORAVEL [Fe/H]
values being only 0.074.

In the case of magnetic Ap stars, the situation is complicated in three ways:
\begin{itemize}
\item The abundances are not standard (i.e. solar scaled), but some elements
are considerably enhanced -- especially the rare earths in cool Ap stars
considered here -- while others are underabundant (e.g. He, C, N, O) in the
atmosphere. Therefore, one can no more, in principle, consider [Fe/H] as a
meaningful metallicity indicator. Nevertheless, in practice $W$ should remain
strongly correlated with [Fe/H], since most lines of CORAVEL's mask are
neutral iron lines, or at least lines of iron-peak elements.
\item The abundances are not distributed uniformly over the stellar surface,
but are often concentrated in patches whose positions depend on the geometry
of the magnetic field. This implies that potentially, all three parameters
of the correlation dip may vary as a function of time, according to the
rotation of the star. Significant changes of the width and depth of the dip
are indeed observed in some cases, but remain often negligible. Radial
velocity variations due to abundance patches or spots sometimes occur as well,
when the star has a significant $v\sin i$, since spots with enhanced lines of
Fe-peak elements will contribute most to the correlation dip.
Thus, in some favourable cases, one can determine the rotational periods.
\item The magnetic field may widen the dip through the Zeeman effect and also
enhance its equivalent width through Zeeman intensification of the lines.
As a result, those stars which have a strong surface field $H_{\mbox{s}}$
(the disk-averaged modulus of the field, see e.g. Preston \cite{P71}) yield so
large a dip that they mimick normal stars with a much larger $v\sin i$.
Therefore, the $v\sin i$ estimates based on the calibration of Benz \& Mayor
(\cite{BM81}, \cite{BM84}) can only represent, in general for the magnetic Ap stars, an
{\it upper limit} to their true $v\sin i$. There is no mean of disentangling
$v\sin i$ and $H_{\mbox{s}}$ in such cases, unless one of these quantities is
determined independantly. Though this is true in the case of CORAVEL, such
disentangling is possible with the more efficient ELODIE spectrograph when two
different masks are used (one selecting lines sensitive to Zeeman effect, the
other selecting lines less sensitive to it), as shown by Babel et al.
(\cite{BNQ95}) and by Babel \& North (\cite{BN97}).
\end{itemize}

\subsection{Zeeman effect}
The latter complication is illustrated in Fig. 1, where the dip
width $\sigma$ is plotted as a function of $v\sin i$, which was taken from
the literature, first from Preston (\cite{P71}), then from Abt \& Morrell
(\cite{AM95})
and from Renson (\cite{R91}). Since the interesting range of $v\sin i$ values is
between 0 and 30 km\,s$^{-1}$ and the relative errors are often very large
(depending on the spectral resolution, the values given in the literature
are sometimes only upper limits), we have taken the minimum value among these
three sources, when the star was mentioned in more than one of them.
Furthermore, we have estimated the equatorial velocity from the rotational
period (given by Renson \cite{R91} or updated by Catalano et al. \cite{CRL93} and by
Catalano \& Renson \cite{CR97}) and from the radius through the oblique rotator
formula $v_{\mbox{eq}}=50.6\times R/P_{\mbox{rot}}$ (with $v_{\mbox{eq}}$ expressed in
km\,s$^{-1}$, $R$ in solar radii and $P_{\mbox{rot}}$ in days, see Stibbs
\cite{s50}), and substituted the measured
$v\sin i$ value by $v_{\mbox{eq}}$ whenever the latter was smaller than the
former. The radius was obtained from the Hipparcos parallax and a photometric
estimate of the effective temperature, in the way
described by North (\cite{N98}). The quantity $v_{\mbox{eq}}$ remains an upper limit
to the true $v\sin i$ because of the projection effect, but in some cases
it is tighter than the observational upper limit.
\begin{figure}
\resizebox{\hsize}{!}{\includegraphics[width=9.0cm]{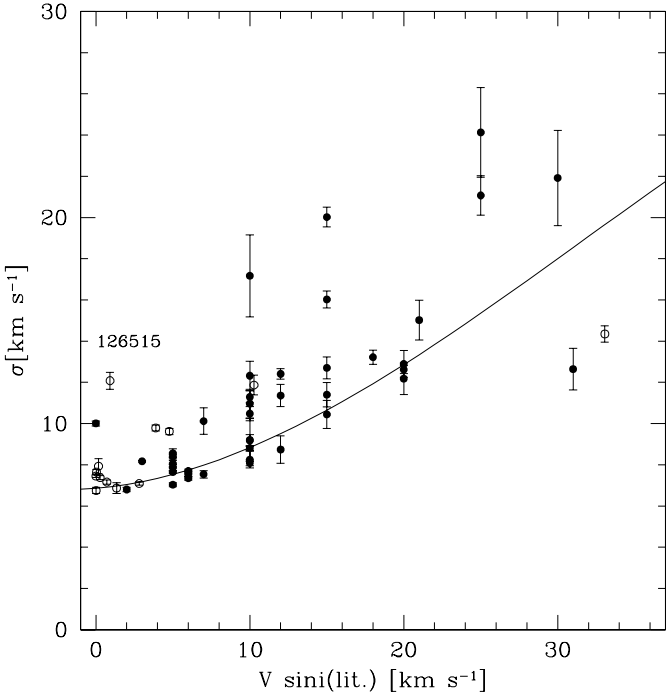}}
\caption[]{Width of the gaussian fitted to the CORAVEL correlation dip
as a function of $v\sin i$ of some well studied Ap stars.
Full dots: observed $v\sin i$ from the literature.
Open dots: $v_{\mbox{eq}}$ computed from the rotational period and estimated
radius through the oblique rotator formula, when it was smaller than the
published $v\sin i$ (see text).
The full line is the
calibration of Benz \& Mayor (\cite{BM84}) for normal F--G stars, extrapolated for
effective temperatures relevant to Ap SrCrEu stars (around 8000~K). Notice the 
discrepant position of some strongly magnetic stars like HD 126515.}
\label{vsini}
\end{figure}
\begin{figure}
\resizebox{\hsize}{!}{\includegraphics[width=8.8cm]{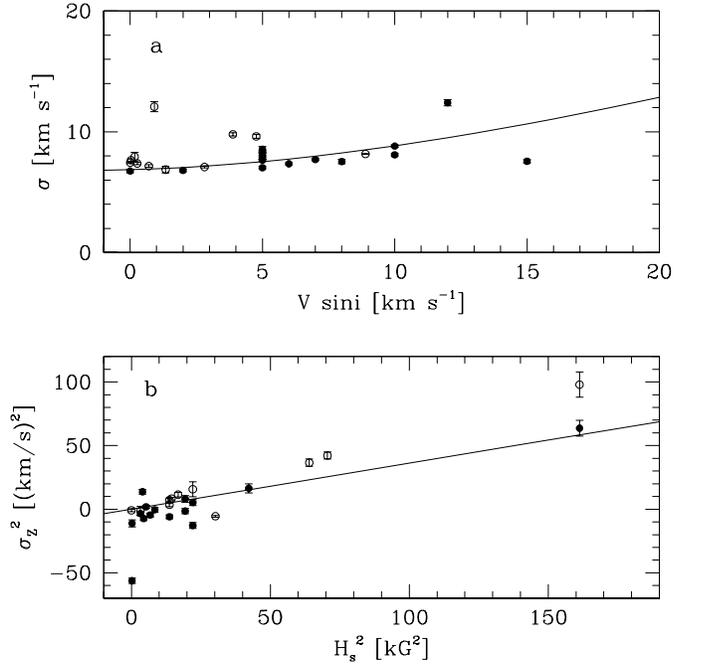}}
\caption{{\bf a:} Same as Fig. 1, but only for stars with a known surface
magnetic field.{\bf b:} Quadratic difference between the observed dip width
and the value expected from the observed $v\sin i$, using the calibration
of Benz \& Mayor (\cite{BM81}, \cite{BM84}), versus $\langle H_s\rangle^2$. The line is the
theoretical relation between $\sigma_{\mbox{Z}}^2$ and $\langle H^2\rangle $,
assuming $\langle g_{\mbox{eff}}\rangle =1.0$ (Eq. 5).}
\label{sigma}
\end{figure}
In Fig. 1, the continuous line is the calibration of Benz \& Mayor (\cite{BM81}, \cite{BM84}) 
for $T_{\mbox{eff}}=8000$~K, i.e. for $B-V=0.183$ according to Equ. (4) of
Hauck (\cite{H85}); it is almost identical to the curve shown in Fig. 3
of Benz \& Mayor (\cite{BM81}).
Although many Ap stars fit the normal relation fairly well, some lie
significantly above it. Especially conspicuous is the case of HD 126515, which
has a well-known rotational period of 130 days (e.g. Mathys et al. \cite{MHL97}) and a
strong surface magnetic field which varies between 10 and 17 kG
(Preston \cite{P71}, Mathys et al. \cite{MHL97}): the large $\sigma$ can
evidently not be attributed to $v\sin i$ in this case!


Figure 2a is the same as Fig. 1, but restricted to stars with a known surface
magnetic field, which naturally restricts the sample to slow rotators.
The surface field has been taken from Mathys et al. (\cite{MHL97}) and from
Preston (\cite{P71}).
Figure 2b shows the relation between $\sigma_Z^2 = \sigma^2-\sigma_V^2(v\sin i)$
and the square of the surface field,
where $\sigma_V^2(v\sin i)$ is the $\sigma^2$ of the dip expected from the
$v\sin i$
through the calibration of Benz \& Mayor and $\sigma$ is the measured, total
width. Clearly, a significant correlation emerges, confirming the sensitivity
of the correlation dip to Zeeman effect. The points follow roughly
the theoretical relation shown by the line,
which is determined in the following way.

The wavelength shift $\Delta\lambda$ between the centre of gravity of the
$\sigma_\pm$ component of a normal Zeeman triplet, and the central
wavelength $\lambda_0$ of the line without magnetic field is given by
\begin{equation}
\Delta\lambda = 4.67\times 10^{-13}\lambda_0^2\, g_{\mbox{eff}} \langle H\rangle 
\end{equation}
where $\lambda_0$ is in \AA, $g_{\mbox{eff}}$ is the effective Land\'e factor
and $\langle H\rangle $, in Gauss, is the modulus of the magnetic field averaged over the
visible stellar disk. Dividing this relation by $\lambda_0$ and multiplying it
with the speed of light $c$, one gets the following expression for the Zeeman shift
(expressed in terms of an equivalent Doppler velocity), assuming,
furthermore, that we take a mean over a large number of lines:
\begin{equation}
\langle \Delta v\rangle = 1.40 \times 10^{-7}\lambda_{norm} \langle g_{\mbox{eff}}\rangle\langle H\rangle
\end{equation}
where $\Delta v$ is expressed in km\,s$^{-1}$, $\lambda_{norm}$ is a
normalisation wavelength (chosen e.g. in the middle of the CORAVEL
wavelength range) expressed in \AA, and $\langle g_{\mbox{eff}}\rangle $ is the effective Land\'e
factor averaged over all spectral lines considered (see Babel et al.
\cite{BNQ95}).

Assuming now that all sources of broadening of the spectral lines are small
enough that the resulting profile remains gaussian, one can cumulate them by
a simple quadratic addition:
\begin{equation}
\sigma^2=\sigma_0^2+\sigma_{\mbox{rot}}^2+
(1.40\times 10^{-7})^2\lambda_{\mbox{norm}}^2 \langle g_{\mbox{eff}}^2\rangle \langle H^2\rangle
\end{equation}
where $\sigma_0$ is the sum of the intrinsic and instrumental widths of the
lines of a non-rotating
and non-magnetic star (it includes the thermal width and depends, therefore,
on the effective temperature) and $\sigma_{\mbox{rot}}$ is only due to projected
rotational velocity; both $\sigma_0$ and $\sigma_{\mbox{rot}}$ are in
km\,s$^{-1}$. Here, because the Zeeman components remain unresolved,
the quantities $g_{\mbox{eff}}^2$ and $H^2$ are averaged, 
instead of $g_{\mbox{eff}}$ and $H$ as in Equ. 2. The quantity
$\langle g_{\mbox{eff}}^2\rangle$ may be defined as
\begin{equation}
\langle g_{\mbox{eff}}^2\rangle = \frac{\sum_{i=1}^{N} (d_i \frac{\lambda_{0i}}
{\lambda_{\mbox{norm}}} g_{\mbox{eff i}})^2}{\sum_{i=1}^{N} d_i^2}
\end{equation}
where $d_i$ is the residual depth of the $i_{th}$ spectral line.

In Equ. 3, the quantity $\sigma_0^2+\sigma_{\mbox{rot}}^2$ 
represents the width of the correlation dip expected for a normal, non-magnetic 
rotating star, from the calibration of Benz \& Mayor; the third term, which we
call $\sigma_{\mbox{Z}}^2$,
represents the additional contribution of the Zeeman effect. $\sigma$ is the
observed total width.

In practice, we chose $\lambda_{\mbox{norm}}= 4300$~\AA, roughly corresponding to
the centre of the CORAVEL passband, and assumed $\langle g_{\mbox{eff}}^2\rangle =1.0$;
$\sigma_{\mbox{Z}}$ then becomes:
\begin{equation}
\sigma_{\mbox{Z}}^2=\sigma^2-(\sigma_{\mbox{int}}^2+\sigma_{\mbox{rot}}^2)
= 0.362 \langle H^2\rangle 
\end{equation}
where $H$ is expressed in kG.
This result is in rough agreement with the empirical data shown in Fig. 2

\section{Intrinsic variations of the correlation dip}
By ``intrinsic'', we mean here those variations of the correlation dip which
are due to the presence of abundance patches on the surface of the star and
to the axial rotation. On the one hand, these variations represent a ``noise''
regarding the precise determination of the radial velocity of the star as
a whole; on the other hand, they represent an additional information which,
in some cases, allows an estimate of the rotational period independantly from
e.g. photometry. In any case, such variations draw the attention towards those
stars which are spectrum variables and may serve as targets for detailed
Doppler imaging.

\begin{figure}
\resizebox{\hsize}{!}{\includegraphics[width=8.8cm]{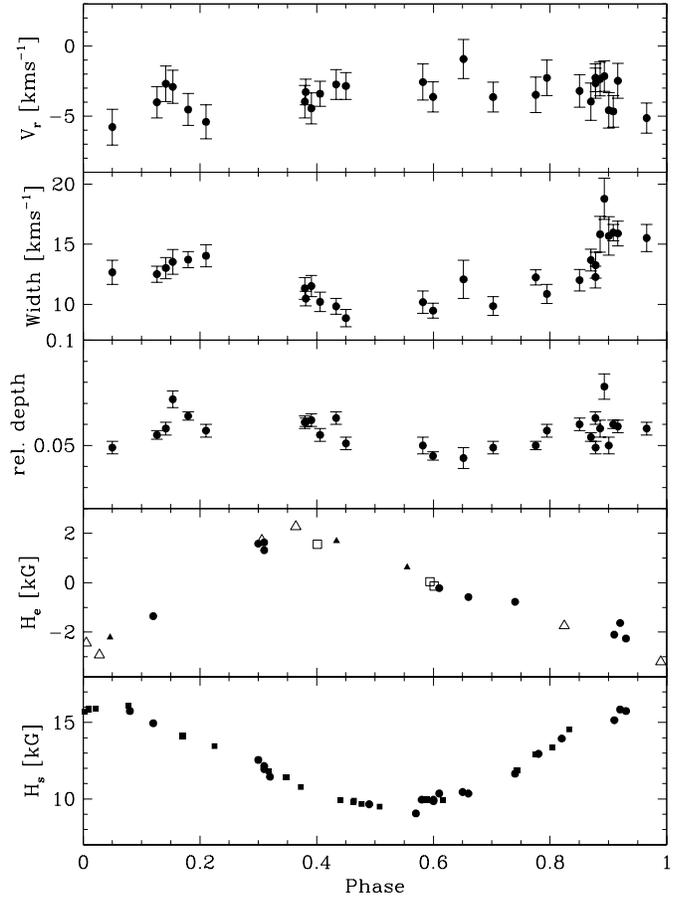}}
\caption{Variations of the CORAVEL correlation dip as a function of
rotational phase for the slowly rotating Ap star HD 126515. These variations
are due to abundance patches. The three upper panels show the moment of first
order ($RV$) of the dip, its width and its depth. The 4th panel shows
the longitudinal magnetic field (key to symbols: black dots,
Preston \cite{P70};
open squares, van den Heuvel \cite{vH71}; open trianges, Mathys \cite{M94};
full triangles, Mathys \& Hubrig \cite{MH97}). The lower panel shows the
surface magnetic field (key to symbols: black dots, Preston \cite{P70}; full
squares, Mathys et al. \cite{MHL97}).}
\label{hd126}
\end{figure}

In Fig.~\ref{hd126} are shown the $RV$, dip width and depth variations of the
well-known spectrum variable HD 126515. The ephemeris is
\begin{equation}
HJD(H_{\mbox{s}} max) = 2437015.000 + 129.95 E,
\end{equation}
the zero point being defined by Preston (\cite{P70}) and the period being
refined
by Mathys et al. (\cite{MHL97}). For this particular star, the shape of the dip
is clearly changing according to the rotational phase, but the radial velocity
remains roughly constant. The rotational velocity is so low that even
contrasted abundance patches will not affect $RV$ much. The relation
between the surface field and the dip variations are not straightforward:
the dip width seems to vary as a double wave, while $H_{\mbox{s}}$ varies as a
single wave. There is a trend for the dip to become wider as $H_{\mbox{s}}$
increases.

\begin{figure}
\resizebox{\hsize}{!}{\includegraphics[width=8.8cm]{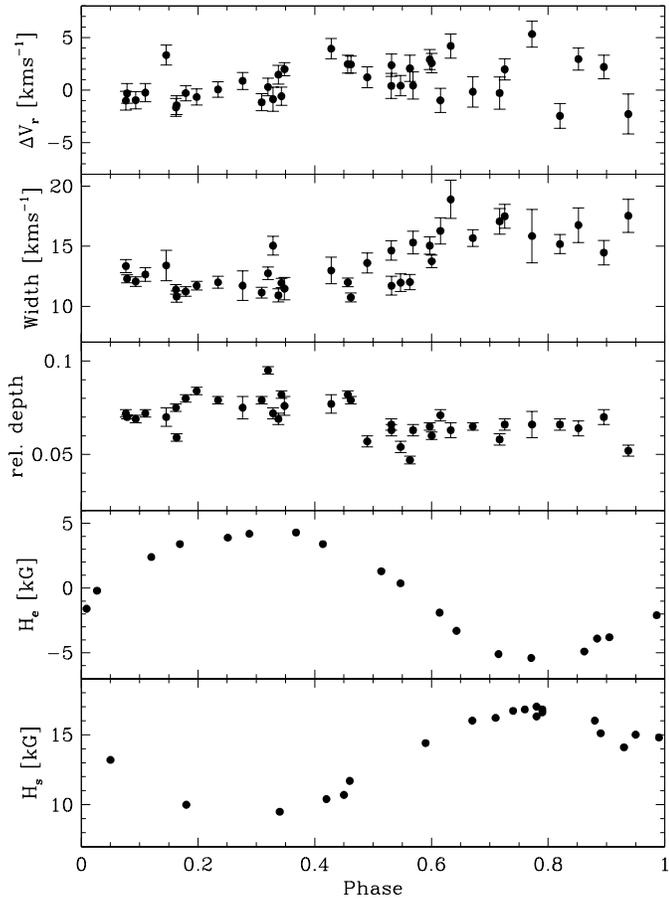}}
\caption{Same as Fig.~\ref{hd126}, but for the Ap star HD 65339.
The longitudinal magnetic field is from Borra \& Landstreet (\cite{BL77}).
The surface magnetic field is from Huchra (\cite{H72}).}
\label{hd65}
\end{figure}

In Fig.~\ref{hd65} are shown the same quantities for the well known Ap star
HD 65339 (or 53 Cam), though the fitted orbital radial velocities (see next
Section) have been subtracted to the observed ones to obtain the variations
$\Delta V_{\mbox{r}}$ due to rotation and abundance patches only.
The ephemeris used is from Borra \& Landstreet (\cite{BL77}):
\begin{equation}
HJD(pos. cross.) = 2435855.652 + 8.0267 E .
\end{equation}

Here again, the correlation dip appears wider at phases of larger surface
magnetic field, as expected from Eq. 6.
There are also clear ``intrinsic''
$RV$ variations in this star, which raises the question of how
to discriminate between intrinsic and orbital variations. Such a question is
especially acute when trying to estimate the rate of binaries, and was already
tackled by Abt \& Snowden (\cite{AS73}). The solution these authors proposed was to
use the hydrogen lines instead of the metallic ones: hydrogen being largely
predominent, its abundance may be considered homogeneous over the whole
surface of the star. However, the CORAVEL mask precisely excludes all
hydrogen lines, so this solution cannot hold for us. Nevertheless, one
observes that those stars presenting an intrinsic $RV$ variation also
have broad and shallow dips betraying a relatively fast axial rotation
(typically 20 km\,s$^{-1}$). These variations occur on a relatively short
timescale, while their amplitude is rather small: this leads to an extremely
small mass function, if they are interpreted as due to orbital motion. They
might present a problem, in principle, only in the ideal case of an extremely
small and strongly contrasted abundance spot, which would yield a sharp,
narrow dip with a significant $RV$ variation.

An interesting case which illustrates well the above considerations,
is that of the Ap Sr star HD 49976 (HR 2534). It is a conspicuous
spectroscopic variable
which was studied by Pilachowski et al. (\cite{PB74}). These authors found a
rotational period $P=2.976$~days, which was later refined to
$P=2.97666\pm 0.00008$~days by Catalano \& Leone (\cite{CL94}) by means of $uvby$
photometry. Fig. 5, based on the ephemeris
\begin{equation}
HJD(pos. cross.) = 2441298.76 + 2.97666 E
\end{equation}
shows a rather well defined, double wave $RV$ 
variation whose peak-to-peak amplitude is as large as 15 km\,s$^{-1}$. This is
consistent with two spots. There
is one maximum around phase 0.3 and another one around phase 0.8. This is
qualitatively consistent with the $RV$ variation of the Fe and Cr
lines obtained by Pilachowski et al. (\cite{PB74}, their Fig. 1), though the maxima
fall rather near phases 0.4 and 0.9 in their case. The width and depth of the
dip are ill-defined because of the large rotational velocity
($31\pm 3$~km\,s$^{-1}$ according to these authors).

\begin{figure}
\resizebox{\hsize}{!}{\includegraphics[width=8.8cm]{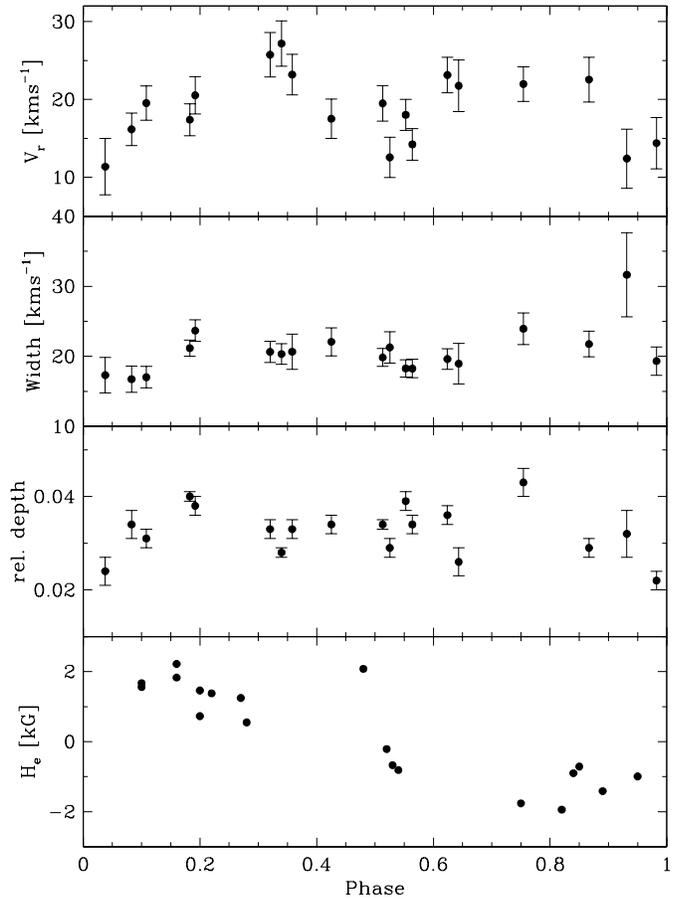}}
\caption{Same as Fig. 3, but for the Ap star HD 49976.
The longitudinal magnetic field is from Pilachowski et al. (\cite{PB74}).}
\label{hd49}
\end{figure}

Finally, we give in Fig. 6 the periodogram of the variation of the dip
depth of HD~5797 (V551 Cas), to illustrate how such variations can help
to determine rotational periods. We obtain $P_{\mbox{rot}}=68.02\pm 0.10$ days, 
compatible with one of the possible periods proposed by Wolff (\cite{W75}).

\begin{figure}
\resizebox{\hsize}{!}{\includegraphics[width=9.2cm]{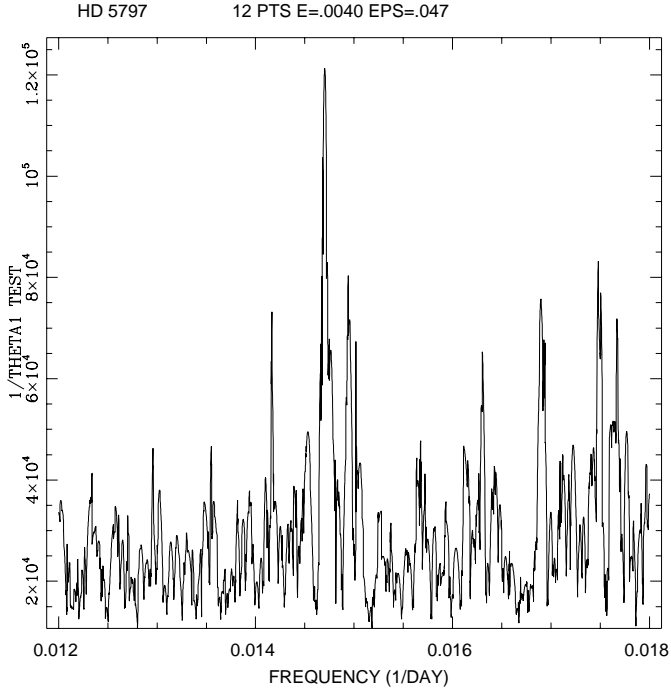}}
\caption{Periodogram of the variation of the depth of the correlation
dip for the Ap CrEuSr star HD~5797. Although it is based on only 12 
observations, the peak is well defined and unique in the whole range of
frequencies from 0 to 0.03 days$^{-1}$.}
\label{period}
\end{figure}
The periodogram is based on the method of Renson (\cite{R78}), but the reciprocal
of his $\theta_1$ test has been plotted. There are only twelve observations
spanning 5873 days, but it is interesting to see that the other periods
suggested by Wolff (\cite{W75}) as acceptable from her photometric data, 45.5
and 57 days, are clearly rejected in view of our data, even though the latter
are scattered on a so long time range. Interestingly,
Barzova \& Iliev (\cite{BI88}) and Iliev et al. (\cite{I92}) have also rejected
the two shorter periods on the basis of high resolution spectra, but admit both
67.5 and 69 days as equally possible.

We have verified that our period agrees with the photometric data of
Wolff (\cite{W75}): Renson's method applied to the differential $b$ magnitudes,
which seem to have the best S/N ratio, gives $P_{\mbox{rot}}=68.2\pm 1.0$~days.
Likewise, the $v$ magnitudes, which also have a good S/N, give
$P_{\mbox{rot}}=67.6\pm 0.5$~days. The $y$ and $u$ magnitudes, which are more
noisy (at least compared to the amplitude of variation) give a best period
around 57 days, but are also compatible with the 68 days period.
Therefore, our 68.02 days period is quite compatible with existing data and
is probably the best estimate available to date.
\begin{figure}
\resizebox{\hsize}{!}{\includegraphics[width=8.8cm]{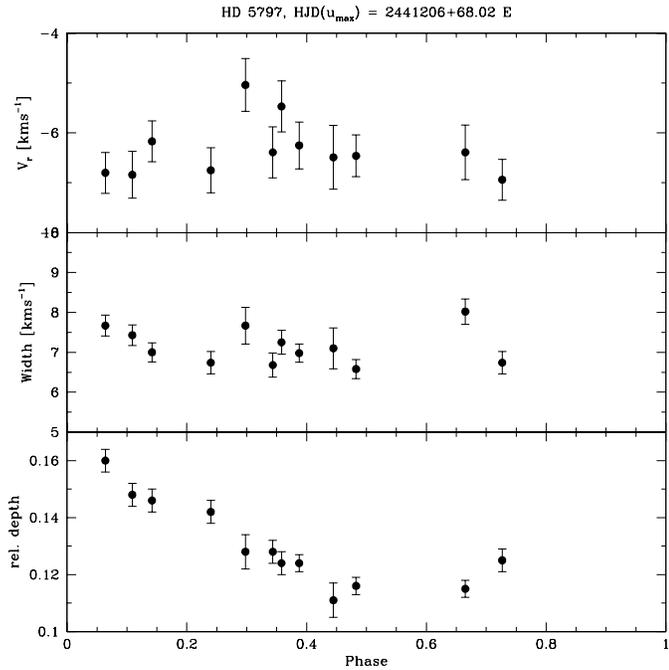}}
\caption{Same as Fig. 5, but for HD~5797. No magnetic field curve exists for
this star.}
\label{hd57}
\end{figure}
The variation of the three parameters of the correlation dip of HD~5797 is
shown on Fig.~7. Unfortunately, the phase coverage is not good, but the depth
of the dip varies in a very significant way, contrary to the radial velocity
and dip width. No longitudinal field curve exist for this
star, although Preston (\cite{P71}) estimated a surface field of 1.8 kG.

\section{Statistics of Ap binaries}

\subsection{The sample of binaries}

Our sample is composed of all Ap stars known as spectroscopic binaries. The 
catalogue of Renson (\cite{R91}) gives us the Ap stars. However, some stars
considered as Am by Renson, have been considered in this work as Ap (HD~56495,
HD~73709 and HD~188854).
Among these stars, the spectroscopic binaries have been selected from the 
following sources. Fourteen orbits were determined
thanks to the CORAVEL scanner (this paper and North et al. \cite{NC98}).
Lloyd et al. (\cite{LS95}) determined a new orbit for $\theta \ Carinae$ (HD~93030) and
Stickland et al. (\cite{st94}) for HD 49798. Recently, Leone \& Catanzaro (\cite{LC99})
have published the orbital elements of 7 additional CP stars, two of which are
He-strong, two are He-weak, one is Ap Si, one Ap HgPt and one Ap SrCrEu.
Wade et al. (\cite{WD00}) provided the orbital elements of HD 81009.
Other data come from the Batten et al. (\cite{BFM89}) and
Renson (\cite{R91}) catalogues. Among all Ap, we kept 78 stars with known
period and eccentricity, 74 of them having a published mass function.

\subsection{Eccentricities and periods}
\begin{figure}
\resizebox{\hsize}{!}{\includegraphics[width=9.0cm]{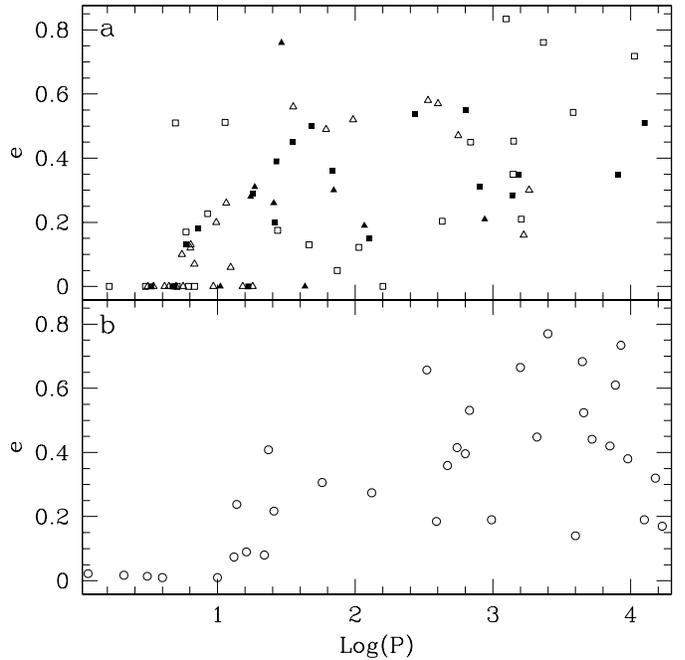}}
\caption{{\bf a:} Diagram eccentricity versus period (in days) for the 78 Ap stars.
The symbols are according to the type of Ap: $\triangle$
HgMn, $\blacktriangle$ He~weak, $\square$ SrCrEu, $\blacksquare$ Si.
{\bf b:} Diagram eccentricity versus period (in days) for G dwarf stars
(Duquennoy \& Mayor \cite{DM91}).}
\label{elogp}
\end{figure}

The statistical test of Lucy \& Sweeney (\cite{lucy}) has been applied to all
binaries with moderate eccentricity, in order to see how far the latter is
significant. The eccentricity was put to zero whenever it was found
insignificant.
We can notice the effect of tidal interactions (Zahn \cite{Z77}, \cite{Z89},
Zahn \& Bouchet \cite{ZB89}) on the orbits of Ap stars (see Fig.~\ref{elogp}a).
Indeed, all orbits with $P$ less than a given value ($=P_{circ}$) are
circularized. According to the third Kepler law, a short period implies a
small orbit where tidal forces are strong.
The period-eccentricity diagram for the Ap stars does not show a well marked
transition from circular to eccentric orbits, in the sense that circular as well
as eccentric systems exist in the whole range of orbital periods between
$P_{circ}=5$~days to a maximum of about 160 days. The wider circular orbits
probably result from systems where the more massive companion once went through
the red giant phase; the radius of the former primary was then large enough
to circularize the orbit in a very short time. This is quite consistent with the
synchronization limit for giant stars of 3 to 4 $M_{\odot}$($P\sim 150$~days,
Mermilliod \& Mayor \cite{MM96}).

An upper envelope seems well-defined in the $e$ vs. $\log P$ diagram,
especially for $\log P < \sim 2$, although four points lie above it.
The leftmost of these, with ($\log P$,$e$) = (0.69,0.52) has a rather
ill-defined orbit\footnote{21 Her (HD~147869) has been measured by Harper
(\cite{H31}) and has a relatively small amplitude.}. Whether this envelope has
any significance remains to be confirmed with a larger sample than presently
available.

It is interesting to compare our $e/\log P$ diagram with that established by
Debernardi (\cite{DebThesis}) for his large sample of Am stars. For these stars,
the limit between circular and eccentric orbits is much steeper, many systems
having $e\sim 0.7$ at $P=10$~d only. The difference might be due to the wider
mass distribution of the Bp-Ap stars (roughly 2 to 5 $M_\odot$) compared to that
of Am stars (1.5 to 2-2.5 $M_\odot$),$P_{circ}$ being different for each mass.
One might also speculate that the lack of very eccentric and short periods is
linked with the formation process of Ap stars, which for some reason (e.g.
pseudo-synchronization in the PMS phase, leading to excessive equatorial
velocities?) forbid this region.

Qualitatively, Ap stars behave roughly like normal G-dwarfs (Duquennoy \& Mayor
\cite{DM91}) in the eccentricity-period diagram. There is also a lack of
low eccentricities at long periods (here for $\log P > 2.0$), and the upper
envelope is similar in both cases for $\log P > 1.0$. One difference is the
presence of moderate eccentricities for periods shorter than 10 days among
Ap binaries, and another is the complete lack of very short orbital periods
($P\leq 3$~days) among them, if one excepts HD 200405. The latter feature,
already mentioned in the past (e.g. by GFH85) is especially striking because
it does not occur for the Am binaries. The physical cause for it is probably
that synchronization will take place rather early and force the components to
rotate too fast to allow magnetic field and/or abundance anomalies to subsist.
However, even an orbital period as short as one day will result in an equatorial
velocity of only 152 km\,s$^{-1}$ for a 3~$R_{\odot}$ star, while single Bp or
Ap stars rotating at this speed or even faster are known to exist (e.g. HD
60435, Bp SiMg, $P_{rot}=0.4755$~d; North et al. \cite{NBL88}). The detection
of one system with $P_{orb}=1.635$~d further complicates the problem, even
though it remains an exceptional case as yet.

The shorter $P_{circ}$ value can be explained by the following effects:
\begin{itemize}
\item Ap stars are younger on average than G dwarfs, leaving less time
for circularization
\item Ap stars have essentially radiative envelopes, on which tidal effects
are less efficient: the circularization time increases much faster with the
relative radius than for stars with convective envelopes: $t_{circ}
\propto (a/R)^{21/2}$ instead of $t_{circ}\propto (a/R)^8$ (see e.g.
Zahn \cite{Z77}), $a$ being the semi-major axis of the relative orbit and $R$
the stellar radius.
\item Therefore, it is essentially the secondary component which
is responsible for the circularization of systems hosting an Ap star. Its radius
being smaller than that of the Ap star, the circularization time is longer.
Thus $=P_{circ}$ is shorter in these systems than in those hosting G dwarfs.
\end{itemize}
Our results confirm those of GFH85. Many stars of
the literature have imprecise orbital elements, making a detailed analysis
difficult. We have not enough CORAVEL orbits of Ap
stars yet to make a more precise statistics.

\subsection{Mass function}

\begin{figure}
\resizebox{\hsize}{!}{\includegraphics[width=9.0cm]{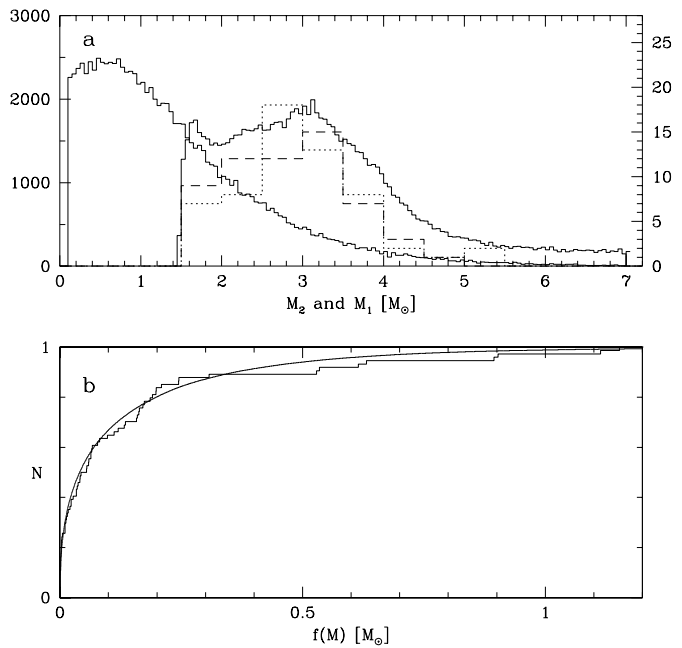}}
\caption{{\bf a.} Distributions of the masses of the secondary (left) and of
the primary (right) used in simulation (100000 dots). The dotted and dashed lines represent 
the distributions of the primary masses of 60 stars of the sample respectively with and without applying the Lutz-Kelker correction
(\cite{lk73}). {\bf b.} Observed and
simulated cumulative distribution of the mass functions (including the He-weak
and He-strong stars).}
\label{cumul}
\end{figure}

The mass function contains the unknown orbital inclination $i$.
Therefore, neither the individual masses
nor the mass ratio can be calculated from it. However the orbital inclination of
Ap stars can be assumed to be randomly oriented on the sky.
Thus, we can compare the observed cumulative distribution of the mass functions
(for our 74 stars) with a simulated distribution (see Fig.~\ref{cumul}), where
we assume random orbit orientations.
We approximated the observed relative distribution of Ap masses
(North~\cite{N93}) by the function given below, then multiplied it by Salpeter's
law (${\cal M}^{-2.35}$) in order
to obtain the simulated distribution of the primary masses $f_{Ap}$:
\begin{equation}
f_{Ap} = [ 0.3623 + 0.8764 \cdot {\cal M} + 8.481 \cdot e^{-({\cal M} - 3.5)^2}]
\cdot {\cal M}^{-2.35}
\end{equation}
with a minimum mass of $1.5 {\cal M_\odot}$ and a maximum mass of
$7 {\cal M_\odot}$.

In order to check the simulated distribution of masses of primaries, we
estimated directly the masses of 60 primaries in systems having Geneva and
$uvby$ photometric measurements, as well as Hipparcos parallaxes. The estimation
was done by interpolation through the evolutionary tracks of Schaller et al.
(\cite{SS92}), using the method described by North (\cite{N98}). The mass
estimate was complicated by the binary nature of the stars under study. In the
case of SB2 systems, the $V$ magnitude was increased by up to 0.75, depending on
the luminosity ratio, and the adopted effective temperature was either the
photometric value or a published spectroscopic value. In the case of SB1
systems, only a fixed, statistical correction $\Delta V=0.2$ was brought to the
$V$ magnitude, following North et al. (\cite{NJ97}), which is valid for a
magnitude difference of 1.7 between the components. The effective temperatures
corresponding to the observed colours
were increased by 3.5\%, to take into account a cooler companion with the same
magnitude difference (assuming both companions are on the main sequence). In
general $T_{\rm eff}$ was obtained from Geneva photometry, using the
reddening-free $X$ and $Y$ parameters for stars hotter than about 9700~K and the
dereddened (B2-G) index for cooler stars. The recipe used is described by North
(\cite{NIAU98}). Existing $uvby\beta$ photometry was used to check the validity
of the $T_{\rm eff}$ estimates and the $E(b-y)$ colour excess was used to
estimate roughly the visual absorption through the relation $A_V=4.3 E(b-y)$.
The latter may be underestimated in some cases because the $b-y$ index of Ap
stars is bluer than that of normal stars with same effective temperature, but
only exceptionally by more than about 0.1 magnitude. Reddening maps by
Lucke (\cite{L78}) where used for consistency checks.

The resulting distribution of masses is represented, with an appropriate
scaling, on Fig~\ref{cumul}a, together with the simulated distribution of
primary masses. The agreement is reasonably good, justifying the expression
used above for the distribution of primary masses.

For the distribution of secondary masses, we use the
distribution of Duquennoy \& Mayor (\cite{DM91}) for the mass ratio
$q=\frac{{\cal M}_{2}}{{\cal M}_{1}}$ of nearby G-dwarfs. We assume that the 
companions of the Ap stars are normal. The distribution of the mass ratio is a
gaussian:
\begin{equation}
\xi(q) \propto e^{-\frac{(q-\mu)^2}{2\sigma_{q}^2}}
\end{equation}
with $\mu = 0.23$ and $\sigma_q = 0.42$.
Knowing the mass of the primary and the mass ratio, we can determine the mass of
the secondary. We assume a minimal companion mass of $0.08 {\cal M}_\odot$.
Using the test of Kolmogorov-Smirnov (Breiman \cite{B73}), we find a
confidence-level for the observed and simulated cumulative distributions of
$92 \%
$. After elimination of the He-strong and He-weak stars, the KS test gives
a confidence-level of $98 \%$. In other words, there is only a $2 \%
$
probability that the observed distribution differs from the simulated one.
This indicates that spectroscopic binaries with an Ap primary have the
same distribution of mass ratios as binaries with normal components
(Duquennoy \& Mayor \cite{DM91}).

\subsection{Percentage of cool Ap stars as members of spectroscopic binary
systems}

The sample used for estimating the binary percentage among cool Ap stars is only
composed of the CORAVEL programme Ap stars described in the observation section,
namely 119 stars. Including the stars HD 59435, HD 81009 and HD 137909 which
have been published elsewhere and are not in Tables 1 and 2, the total number
of programme stars is 122. However, 6 of them have only one measurement, so
they are not relevant here; in addition, about 3 stars have very large errors
on their $RV$ values ($\sigma > 3$~\kms), so they are not reliable. Thus there
are 113 objects on which statistics can be done. We chose, as variability
criterion, the probability
$P(\chi^2)$ that the variations of velocity are only due to the internal
dispersion. A star will be considered as double or intrinsically variable
if $P(\chi^2)$ is less than 0.01 (Duquennoy \& Mayor \cite{DM91}). However,
this test can not say anything about the nature of the variability.

For fast rotators, it is difficult to know whether the observed dispersion
is just due to spots or betrays an orbital motion. Among the
sample, 34 stars are assumed to be binaries, namely 30 \%. Nevertheless, this
figure has to be corrected for detection biases; we have attempted to estimate
the rate of detection through a simulation. For this, a
sample of 1000 double stars was created. The mass distributions explained
above are used again. The orbital elements
$T_{\mathrm{o}}$, $\omega$ and $i$ are selected from uniform distributions, while the
eccentricity is fixed to zero for period less than 8 days and is distributed following
a gaussian with a mean  equal to 0.31 and $\sigma=0.04$ (cases with negative
eccentricity were dropped and replaced) for periods less than 1000 days and
larger than 8 days,
and following a distribution $f(e) = 2 e$ for longer periods. The period
is distributed according to a gaussian distribution with a mean equal to
$\overline{log(P)}=4.8$ and $\sigma_{\overline{log(P)}} = 2.3$
(Duquennoy \& Mayor \cite{DM91}), where $P$ is given in days. A cutoff at 2 and
5000 days was
imposed. In a second step, the radial velocities of the created sample are
computed at the epochs of observation of the real programme stars and with
the real errors. Finally the $P(\chi^2)$ value of each star is computed.
We obtain a simulated detection rate of 69 \%.

After correction, we find a rate of binaries among Ap stars of about 43 \%
in excellent agreement with the one determined by GFH85 for the cool Ap stars
of 44 \%. Our sample is however more homogeneous and reliable.

\section{Conclusion}

We have determined a dozen of new spectroscopic orbits of Ap stars. Moreover,
as shown in the Appendix, we
have confirmed the 273-day period of HD~9996 and found a new value for the
period of HD~216533 (P~=~1414.73 days) in complete disagreement with the old
values (16 days). We have computed the mass of both
components of the Ap star $53$~Cam, thanks to our homogeneous radial
velocities which could be combined with the published speckle orbit. We have
also shown that no significant apsidal motion has occurred in the HD 98088
system for the last 40 years.

The main result of this study is that statistically, the orbital parameters
of Ap stars do not differ from those of normal stars, except for an almost
complete lack of orbital periods shorter than 3 days. This cut-off is
accompanied by a parallel lack of circular and low eccentricity orbits,
the latter being due to the former. But in spite of this general rule, there
is the interesting exception of HD 200405, an SB1 system with
$P_{\mbox{rot}}=1.6$~days. This system would merit further investigation.

It is important to mention that the anomalous eccentricity distribution
found by GFH85 is certainly not an independant fact,
but is tightly linked with the lack of orbital periods shorter than 3 days.
So short periods always correspond to circular orbits; therefore, removing them will
result in an apparent excess of high eccentricities. There is nothing abnormal
about the orbital parameters of binary systems hosting an Ap star, except
for the lack of short periods.

The distribution of the mass ratios of Ap binaries is found to be compatible
with the mass ratios of
normal binaries with smaller masses (G-dwarfs). However, the sample of Ap
stars with well-determined orbits is not sufficient to explore possible
differences
between the distributions of orbital parameters of each of the four categories
of Ap stars on the one hand, and of the normal stars on the other hand.

\appendix
\section{Results for individual systems}

\subsection{HD  5550 (= BD +65\degr 115 = Renson 1470)}
This star had originally not been included in the CORAVEL sample because
its photometric effective temperature exceeds 10000~K. On the other hand, it
was included in the sample measured with ELODIE for the survey of surface
magnetic fields, which allowed us to discover its SB2 nature.
The CORAVEL data show that it is a short period binary, and allow us to obtain
the orbital parameters. The SB2 nature of this star is
not visible with the CORAVEL instrument while it is clear with
ELODIE. In addition to the better resolution of ELODIE,
this might be due to the fact that CORAVEL covers only 
the blue wavelength range, while ELODIE is rather sensitive to the red.
The star visible with CORAVEL is the less massive one (see Fig. A.1),
so if significantly redder, the invisible
companion could only be a red giant. This
is forbidden by the short orbital period: the shortest orbital period for
systems hosting a red giant is about 40 days (Mermilliod, private 
communication). Therefore, the more massive component must be also the hotter
one, and its visibility with ELODIE is
probably linked with the mask used for the correlation, which had been
specifically defined for Ap stars, while the CORAVEL mask was designed from
the spectrum of Arcturus. This raises the interesting possibility that
both companions might be Ap stars (or possibly Am), since otherwise their
metallicity would not have been sufficient to yield correlation dips.
Another favorable circumstance is the rather low inclination of the system,
since for an assumed mass $M_1 = 2.5 \, M_\odot$ (for the most massive
component), the inclination $i = 20.5\degr$ only, so that the projected
rotational velocity is only 35 percent of the equatorial value.
\begin{figure*}[h!]
\centering
 \includegraphics[width=17cm]{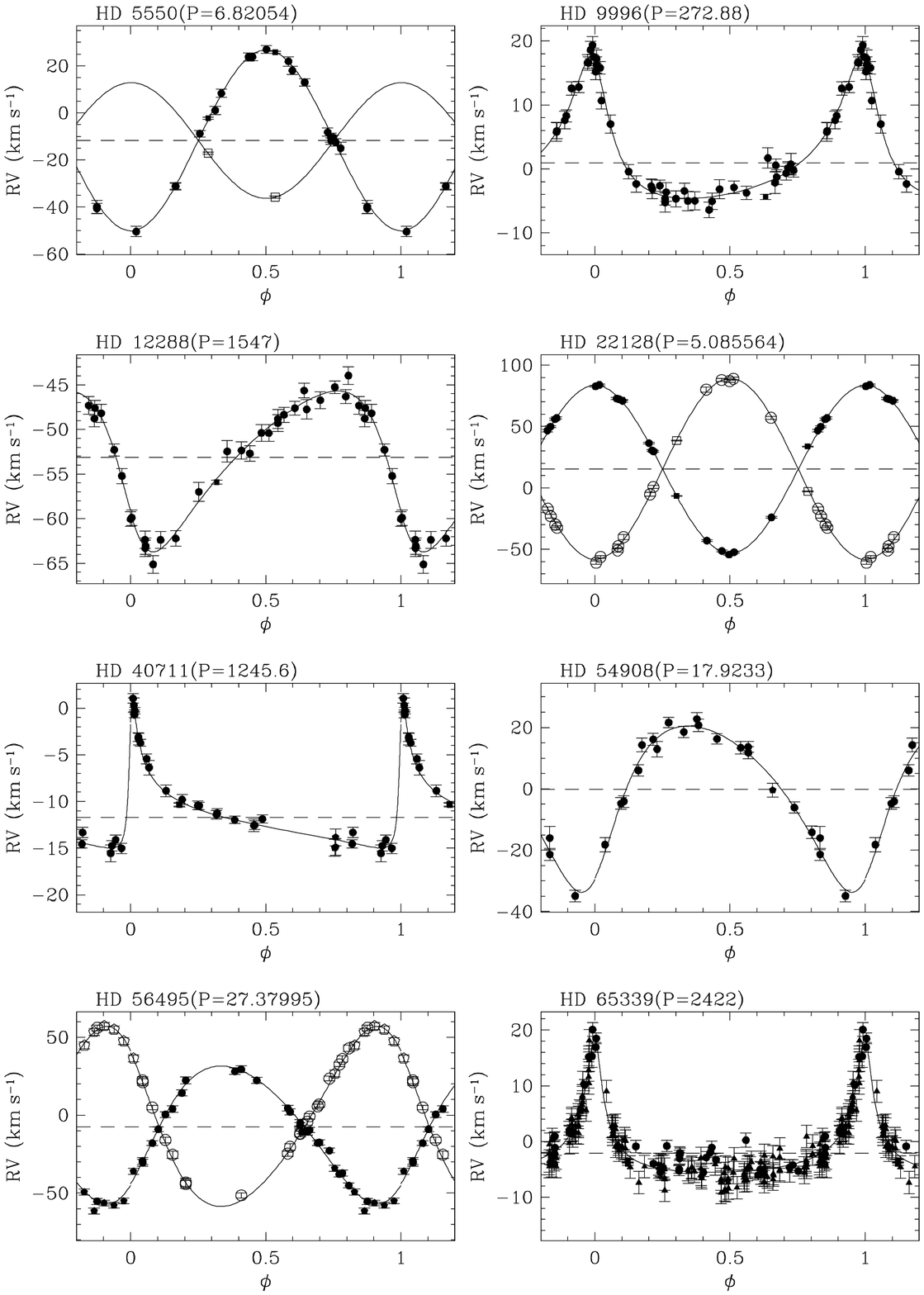}
 \caption{Phase-folded radial-velocity curves of the first eight binaries listed
 in Table 1. Full dots: CORAVEL 
 observations; triangles: ELODIE observations.}
 \label{RVa}
\end{figure*}
\begin{figure*}
\centering
 \includegraphics[width=17cm]{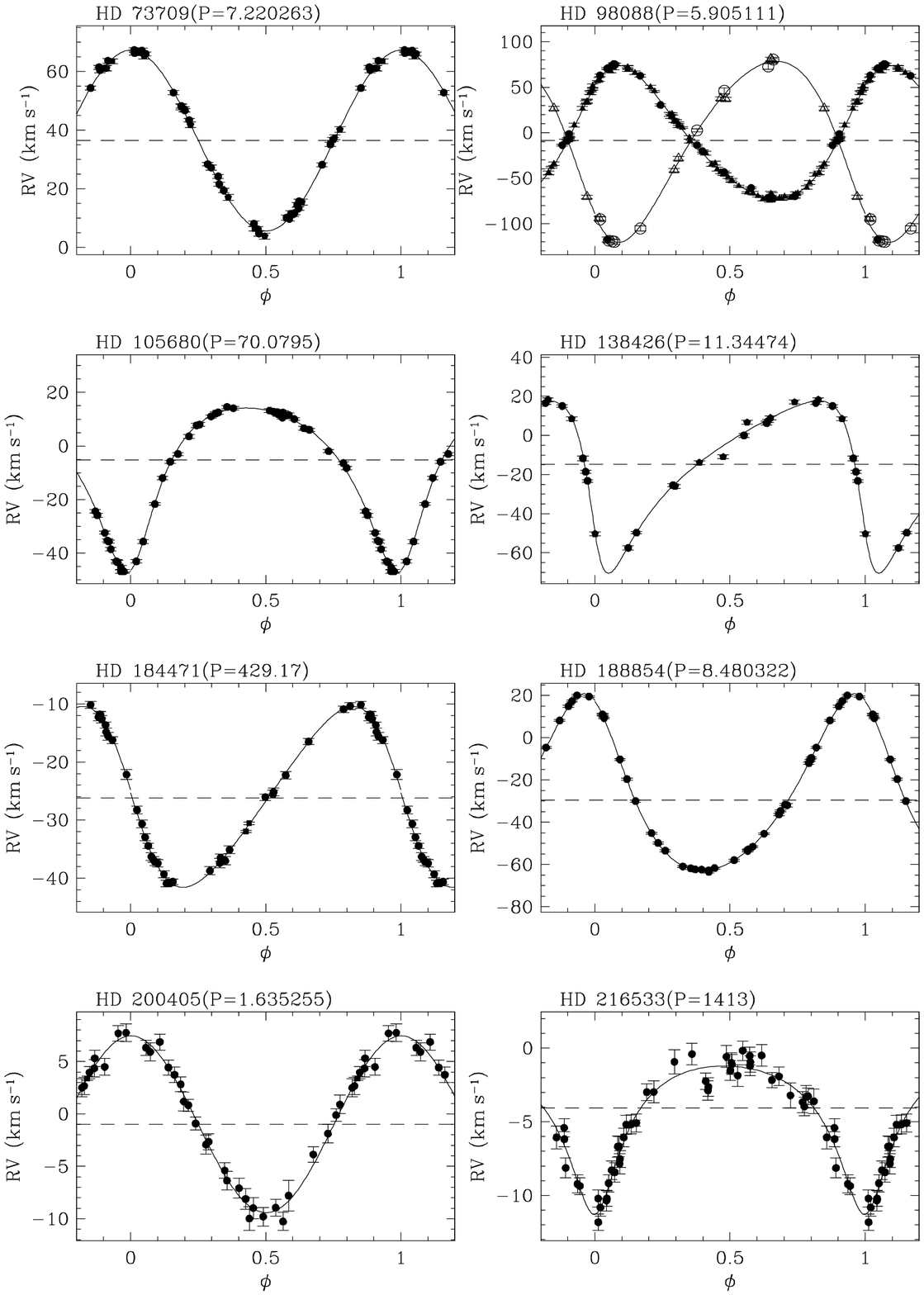}
 \caption{Phase-folded radial-velocity curves of the last eight binaries listed
 in Table 1. Full dots: CORAVEL 
 observations; triangles: ELODIE observations.}
 \label{RVb}
\end{figure*}
The CORAVEL data then do not allow
to obtain the radial-velocity curve of the primary; on the other hand,
they yield the systemic velocity which, together with the two ELODIE
spectra, allow a fairly good estimate of the mass ratio (see Table~\ref{orbel}).

This star is remarkable, because it is only the fifth SB2 system known
among magnetic Ap stars, after HD 55719 (Bonsack \cite{B76}), HD 98088
(Abt et al. \cite{A68}, Wolff \cite{W74}), HD 59435 (Wade et al. \cite{WN96})
and HD 174016-7 (Ginestet et al. \cite{gin99}); HD 59435
had also been studied in the course of this survey, while a sixth
case (HD 22128, see below) was discovered with ELODIE. Although HgMn stars
are frequently seen in SB2 systems, this is exceptional among Si and SrCrEu
stars.

\subsection{HD 9996 (= BD +44\degr 341 = Renson 2470)}

The duplicity of this B9 CrEuSi (Osawa \cite{O65}) star has first been detected
by Preston \& Wolff (\cite{PW70}) who found  an orbital period of 273 days.
They did not attempt to determine the orbital elements because of the poverty
of the data. Scholz (\cite{S78}) tried to determine the orbital elements, but
the shape of the velocity curve in the vicinity of the maximum remained
ill-defined. The 43 CORAVEL measurements (Table~1) confirm the 273-day period (see
Fig.~\ref{RVa}). Thanks to the precision and homogeneity of the data, our
velocity curve is more precise than the one based on the data gathered by
Preston \& Wolff (\cite{PW70}) and Scholz (\cite{S78}), so that a satisfactory
determination of the orbital elements is possible.

The rather large rms scatter of the residuals (1.24 \kms) is due to the small depth of the 
correlation dip (3 percent).
Rotation is not important, the $v\sin i$ value of the visible component is very small
($< 2$ \kms, see Tables 2 and \ref{parphys}) and has no effect on the correlation-peak width;
this is confirmed by the very long rotational period of the primary 
($\sim$~21~years, Rice \cite{Ri88}), so that even highly contrasted
abundance spots could not distort the radial-velocity curve.

\subsection{HD 12288 (= BD +68\degr 144 = Renson 3130)}

This star was classified A2 CrSi by Osawa (\cite{O65}). Its rotational period,
known from its magnetic variability, is 34.79 days
(Mathys et al. \cite{MHL97}). Thirty-one observations were obtained over an
interval of 5949 days (Table~1), which represents about 4 orbital periods
($P_{orb}=1547$~d). The radial-velocity curve is shown in Fig.~\ref{RVa}.
The projected rotational velocity estimated from the width of
the autocorrelation dip is moderate but significant (Tables 2, \ref{parphys}), but should not
be considered as reliable because the effect of the magnetic field is not
taken into account in this estimate. Since the Zeeman effect will always
widen the dip, the $v\sin i$ values listed in Table 2 and \ref{parphys} must be considered as
upper limits to the true projected rotational velocity. If considered with this
caution in mind, they are very useful.
\setcounter{table}{3}
\begin{table*}
\caption{Orbital parameters of the binaries. For each component, the
second line gives the estimated standard deviations of the
parameters.}
\begin{center}
\begin{tabular}{|r|r|r|r|r|r|r|r|r|r|r|} \hline
\multicolumn{1}{|c|}{Star name} & \multicolumn{1}{c|}{$P$} &
 \multicolumn{1}{c|}{$T_\circ$ (HJD} & \multicolumn{1}{c|}{$e$} &
 \multicolumn{1}{c|}{$V_\circ$} & \multicolumn{1}{c|}{$\omega_1$} &
 \multicolumn{1}{c|}{$K_{1,2}$} & \multicolumn{1}{c|}{${\cal
 M}_{1,2}\sin^3 i$} & \multicolumn{1}{c|}{$a_{1,2}\sin i$} &
 \multicolumn{1}{c|}{$N$} & \multicolumn{1}{c|}{(O$-$C)} \\ &
 \multicolumn{1}{c|}{(days)} & \multicolumn{1}{c|}{$-2400000)$} & &
 \multicolumn{1}{c|}{($\mathrm{km\,s^{-1}}$)} &
 \multicolumn{1}{c|}{($^\circ$)} &
 \multicolumn{1}{c|}{($\mathrm{km\,s^{-1}}$)} & 
 \multicolumn{1}{c|}{$f_{1}$($\cal M$)} &
 \multicolumn{1}{c|}{($10^{6}$~
 km)} & & \multicolumn{1}{c|}{($\mathrm{km\,s^{-1}}$)} \\ \hline
HD 5550   & 6.82054 
 & 50988.70& 0.00 & -11.70& - & 24.60  &  0.1081  &    2.307 &   2  & 1.13\\  
 &0.00020& 0.011& fixed&  0.28  &-  &  0.82  &  0.0045  &    0.077 &   & \\  
         &  &  &  &  &  &  &  &  &  & \\
  &  &  &  &  &-  & 38.43 & 0.0692 & 3.605& 22 & \\  
  &  &  &  &  &-  &  0.46&  0.0036& 0.043  &      &    \\
\hline
HD 9996   & 272.88
 & 44492.34 & 0.532 & 0.97 & 20.17 & 11.12 & 0.0237 & 35.34 & 43 &1.33\\
  & 0.20 & 2.24 & 0.023 & 0.22 & 3.35 & 0.29 & 0.0022 & 1.11 & & \\ \hline
HD 12288  & 1546.99
 & 44480.5 & 0.337 & -53.15 & 120.84 & 9.01 & 0.0982 & 180.5 & 31 &
 0.84\\ & 7.29 & 15.8 & 0.024 & 0.16 & 5.49 & 0.26 & 0.0089 & 5.5 & & \\
\hline
HD 22128  & 5.085564
 & 50116.7656 & 0.00 & 15.30 & - & 68.40 & 0.786 & 4.784 & 20 &
 1.25\\ & 0.000070 & 0.0043 &fixed & 0.21 & - & 0.37 & 0.012 & 0.026 & & \\ 
 & & & & & & & & & & \\ & & & & &- & 73.69 & 0.729 & 5.153 & 18
 & \\ & & & & & - & 0.55 & 0.010 & 0.038 & & \\ \hline
HD 40711 & 1245.6 & 49591.7 & 0.834 & -11.69 & 314.3 & 7.88 & 0.0106 &
74.5 & 31&0.51\\
         &    4.4 &     6.3 & 0.013 &   0.12 &   2.1 & 0.46 & 0.0022 &
 5.1 &   &    \\ \hline
HD 54908 
 & 17.9233
 & 46469.96 & 0.286 & -0.15 & 213.45 & 27.09 & 0.0326 & 6.40 & 21 &
 2.58\\ & 0.0017 & 0.36 & 0.034 & 0.61 & 6.47 & 1.19 & 0.0044 & 0.29 & & \\
\hline
HD 56495  & 27.37995
 & 48978.40 & 0.1651 & -7.57 & 224.7 & 44.30 & 1.641 & 16.45 & 32 &
 2.45\\ & 0.00080 & 0.23 & 0.0097 & 0.35 & 3.2 & 0.74 & 0.055 & 0.27 & & \\ 
 & & & & & & & & & & \\ & & & & & 44.7 & 57.75 & 1.259 & 21.44 & 28
 & \\ & & & & & 3.2 & 0.81 & 0.044 & 0.30 & & \\ \hline
HD 65339  & 2422.04
 & 27723.6 & 0.718 & -2.10 & 5.22 & 12.08 & 0.149 & 280.0 & 181 &
 1.72\\
(RV only)& 2.42 & 14.3 & 0.012 & 0.14 & 1.64 & 0.45 & 0.019 & 11.7 & & \\ \hline
HD 73709  & 7.220263
 & 49996.5352 & 0.00 & 36.51 & - & 30.84 & 0.02200 & 3.062 & 45 &
 0.89\\ & 0.000017 & 0.0093 & fixed & 0.13 & - & 0.19 & 0.00041 & 0.019 & & \\
\hline
HD 98088  & 5.905111
 & 34401.387  & 0.1796&-8.45 &314.46&73.29& 1.733& 5.854& 88 & 2.34 \\
 &0.000004&0.023&0.0039&0.23 &  1.44& 0.36& 0.030& 0.029&    &      \\
 &        &   &       &      &134.46&99.46& 1.277& 7.940& 19 &      \\
 &        &   &       &      &  1.44& 0.90& 0.020& 0.070&    &      \\ \hline
HD 105680  & 70.0795
 & 45991.19 & 0.3798 & -5.13 & 192.6 & 30.75 & 0.1676 & 27.42 & 42 &
 0.76\\ & 0.0087 & 0.38 & 0.0055 & 0.13 & 1.1 & 0.20 & 0.0034 & 0.19 & & \\
\hline
HD 138426 & 11.34474 & 48690.398 & 0.512 & -14.63 & 121.37 & 44.02 &
 0.0636 & 5.90 & 20 & 2.04 \\
          &  0.00029 &     0.052 & 0.020 &   0.52 &   2.81 &  1.46 &
 0.0068 & 0.21 &    &      \\ \hline
HD 184471  & 429.17
 & 46857.06 & 0.2017 & -26.16 & 86.99 & 15.59 & 0.1585 & 90.09 & 36 &
 0.51\\ & 0.42 & 3.01 & 0.0081 & 0.12 & 2.81 & 0.15 & 0.0045 & 0.86 & & \\
\hline  HD 188854  & 8.480322
 & 46394.223 & 0.2262 & -29.61 & 23.84 & 41.76 & 0.05926 & 4.743 & 34 &
 0.40\\ & 0.000025 & 0.014 & 0.0024 & 0.07 & 0.64 & 0.10 & 0.00045 & 0.012& &\\
\hline
HD 200405  & 1.635255
 & 46999.9766 & 0.00 & -0.974 & 0.00 & 8.44 & 0.0001021 & 0.190 & 34 &
 0.63\\ & 0.000006 & 0.0058 & fixed & 0.12 & fixed & 0.18 & 0.0000064& 0.004&&\\
\hline  HD 216533 
 & 1413.1
 & 43752.5 & 0.437 & -4.05 & 182.8 & 5.04 & 0.0137 & 88.1 & 48 &
 0.71\\ & 4.6 & 17.4 & 0.026 & 0.11 & 4.9 & 0.22 & 0.0019 & 4.0 & & \\ \hline
\end{tabular}
\end{center}
\label{orbel}
\end{table*}
\begin{table*}
\caption{Orbital parameters of HD 191654, assuming that the RV variations of this star
is caused by orbital motion rather than by rotating abundance patches on its surface.}
\begin{center}
\begin{tabular}{|r|r|r|r|r|r|r|r|r|r|r|} \hline
\multicolumn{1}{|c|}{Star name} & \multicolumn{1}{c|}{$P$} &
 \multicolumn{1}{c|}{$T_\circ$ (HJD} & \multicolumn{1}{c|}{$e$} &
 \multicolumn{1}{c|}{$V_\circ$} & \multicolumn{1}{c|}{$\omega_1$} &
 \multicolumn{1}{c|}{$K_{1,2}$} & \multicolumn{1}{c|}{${\cal
 M}_{1,2}\sin^3 i$} & \multicolumn{1}{c|}{$a_{1,2}\sin i$} &
 \multicolumn{1}{c|}{$N$} & \multicolumn{1}{c|}{(O$-$C)} \\ &
 \multicolumn{1}{c|}{(days)} & \multicolumn{1}{c|}{$-2400000)$} & &
 \multicolumn{1}{c|}{($\mathrm{km\,s^{-1}}$)} &
 \multicolumn{1}{c|}{($^\circ$)} &
 \multicolumn{1}{c|}{($\mathrm{km\,s^{-1}}$)} & 
 \multicolumn{1}{c|}{$f_{1}$($\cal M$)} &
 \multicolumn{1}{|c|}{$10^{6}$~
 km} & & \multicolumn{1}{c|}{$\mathrm{km\,s^{-1}}$} \\ \hline
HD 191654 & 2121. & 48692. & 0.48 & -15.72 & 88. & 2.11&0.00140&54.0&27&0.91\\
          &   27. &    50. & 0.10 &   0.23 & 17. & 0.24&0.00055& 7.1&  &\\
\hline
\end{tabular}
\end{center}
\label{hd191}
\end{table*}

\subsection{HD 22128 (= BD -07\degr 624 = Renson 5560)}

This A7 SrEuMn star (Renson \cite{R91}) was found to be an SB2 system during the
survey for magnetic fields carried out with ELODIE. We do not have Geneva photometry
for that star, but only Str\"omgren photometry\footnote{$V=7.595$, $b-y=0.228$,
$m_1=0.220$, $c_1=0.672$} given by Olsen (\cite{O83}, \cite{O94}).
The average physical
parameters obtained from the uvby$\beta$ values compiled by Mermilliod et al. 
(\cite{MM97}) (assuming both components
are identical) and using the calibration of
Moon \& Dworetsky (\cite{MD85}) are listed in Table~\ref{par22-56}.
From the physical parameters we obtain a typical mass
$\cal M$ = 1.99 $\pm$ 0.17 $\cal M_\odot$, according to the models of
Schaller et al. (\cite{SS92}). The inclination angle $i$ may be estimated close to
48\degr.

Notice, on Fig.~\ref{RVa}, that the radial-velocity curve is very close to a circular
orbit ($e=0.0145 \pm 0.0116$). Therefore, the test of Lucy \& Sweeney (\cite{lucy})
was applied in order to see whether this small eccentricity is significant or not. 
The probability p is equal to 0.69 in our case, which is much greater than the
limit of 0.05 determined by Lucy \& Sweeney.
Thus this eccentricity of 0.0145 is not significant and is fixed to zero.

\begin{table*}
\caption{Physical parameters of HD~22128 and HD~56495 according to their colours
in the uvby$\beta$ or Geneva photometric system.}
\begin{center}
\begin{tabular}{c|clllccccc} \hline \multicolumn{1}{c|}{Star}
& \multicolumn{1}{c}{Photometry} &
\multicolumn{1}{c}{$T_{\mbox{eff}} [K]$} &
 \multicolumn{1}{c}{$\log g$ [cgs]} & \multicolumn{1}{c}{[M/H]}
& \multicolumn{1}{c}{$\log (L/L_{\odot})$} &
 \multicolumn{1}{c}{$R/R_{\odot}$} & \multicolumn{1}{c}{$M_v$}
& \multicolumn{1}{c}{$M_{bol}$} & 
 \multicolumn{1}{c}{$\Delta m_0$}\\ \hline  HD~22128
 & uvby$\beta$ & 6900 & 3.65 & 0.57 & 0.95 & 2.10 & 2.29 & 2.26 & -0.052\\
\hline
HD~56495 &uvby$\beta$ & 7179 & 4.00 & 0.42 & 0.77 & 1.58 & 2.77 & 2.72& -0.032\\
& Geneva &7044 $\pm$ 56 & 4.26 $\pm$ 0.09 & 0.26 $\pm$ 0.08 & & & & & \\
\hline
\end{tabular}
\end{center}
\label{par22-56}
\end{table*}

\begin{table*}
\caption{Physical parameters of the binaries according to their colours in the
Geneva photometric system (or in the uvby$\beta$ system for HD 22128); in the
case of HD 5550, we have adopted the $T_{\rm eff}$ value estimated from the
H$_\alpha$ profile observed with ELODIE near conjonction. The
reddenings E(B2-G) labeled with an asterisk are determined using the maps of
Lucke (\cite{L78}). $v\sin i$ is obtained by a calibration of the CORAVEL 
correlation-dip
width (Benz \& Mayor \cite{BM84}). The resulting $v\sin i$ is slightly less reliable
than for F and cooler stars which were used for the calibration, because their
effective temperature is larger and the corresponding dependance has to be
extrapolated. However, the main source of uncertainty is due to disregarding
the magnetic field, which implies an overestimate of $v\sin i$. The
longitudinal magnetic field is taken from Babcock (\cite{B58}) or more recent
references. $^*$ Value given by Debernardi et al. (\cite{DM20})}
\begin{center}
\begin{tabular}{rrrrrr} \hline \multicolumn{1}{c}{$HD$}
& \multicolumn{1}{c}{$v \sin i$ upper limits } & \multicolumn{1}{c}{$m_V$}
&\multicolumn{1}{c}{$H_z$} & \multicolumn{1}{c}{$T_{\rm eff}$}
& \multicolumn{1}{c}{$E(B2-G)$}\\ 
\multicolumn{1}{c}{}
& \multicolumn{1}{c}{(\kms)} & \multicolumn{1}{c}{}
&\multicolumn{1}{c}{$(KG)$} & \multicolumn{1}{c}{$(\degr \ K)$}
& \multicolumn{1}{c}{}\\ \hline
5550 & $ 6.5\pm 1.3$ & 5.967 & - & 11000 & 0.046 \\
9996 & $ 2.0$ & 6.379 & -1.2 to 0.3 & 9700 & 0.017 \\
12288 & $ 12.5 \pm 0.4$ & 7.748 & -1.2 to -0.2 & 9378 & 0.175\\
22128A & $ 15.9 \pm 0.3$ & 7.595 & - & 7000 & 0*\\
22128B & $ 16.3 \pm 0.7$ &  & - & & \\
40711  & $ 2.0$ & 8.581 & - & 9328 & 0.192* \\
54908 & $ 55.2 \pm 5.5$ & 7.968 & - & 7483 & 0.031\\
56495A & $ 25.3 \pm 2.5$ & 7.654 & 0.21 to 0.57 & 7044 & 0*\\
56495B & $ 12.5 \pm 2.1$ & 7.654 & - & & \\
65339 & $ 19.4 \pm 0.6$ & 6.031 & -5.4 to 4.2 & 8250 & 0.012\\
73709 & $ 17.3 \pm 0.3^*$ & 7.687 & - & 7831 & 0*\\
98088A & $ 21.1 \pm 2.1$ & 6.42 & 0.48 to 0.94 & 8043 & 0* \\
98088B & $ 15.8 \pm 1.9$ & 7.62 &  & 7532 & 0* \\
105680 & $ 14.1 \pm 0.2$ & 8.060 & - & 7154 & 0*\\
138426 & $ 2.0$ & 8.546 & - & 8694 & 0.142 \\
184471 & $ 2.0$ & 8.980 & - & 8114 & 0.116\\
188854 & $ 9.3 \pm 0.2$ & 7.634 & - & 7005 & 0.069*\\
200405 & $ 9.6 \pm 0.4$ & 8.908 & - & 9624 & 0.101\\
216533 & $5.7 \pm 0.3$ & 7.907 & -0.7 to 0.1 & 9000 & 0.120\\
\hline
\end{tabular}
\end{center}
\label{parphys}
\end{table*}

\subsection{HD 40711 (= BD +10\degr 973 = Renson 10880)}
Bidelman \& McConnell (\cite{BM73}) classified this object Ap SrCrEu. Geneva
photometry clearly confirms the peculiarity with $\Delta (V1-G) = 0.020$
(the photometric data in the \textsc{Geneva} system are collected in the General
Catalogue -- Rufener \cite{Ru88} -- and its up-to-date database -- Burki
\cite{Bu02}). The
radial velocities are represented on Figure A.1. The periastron was observed
again only recently, which allowed a precise estimate of the orbital
period. The eccentricity is high and relatively well defined, though the exact
shape of the $RV$ curve in the vicinity of the periastron remains unknown
because of the 7-weeks gap in the observations. The depth of the dip varies,
while its width only shows rather marginal changes.

\subsection{HD 54908 (= BD -01\degr 1579 = Renson 15000)}

HD~54908 is a poorly studied Ap star classified A0 Si by Bidelman \& McConnell
(\cite{BM73}). In spite of a large $v \sin i = 53.6 \pm 5.34$ \kms ,
the variation of the radial velocity is too large to be caused by spots and
rotation ($K = 27.47 \pm 1.14$ \kms ). However we can see the effect of a large
rotational velocity on the scatter of the residuals. The twenty-one
observations were obtained over an interval of 4084 days. The shape of
the radial-velocity curve (Fig.~\ref{RVa}) is not very well defined in the
vicinity of the minimum, but the period of 17.92 days is quite well determined.

\subsection{HD 56495 (= BD -07\degr 1851 = Renson 15430)}

This star was classified A3p Sr by Bertaud (\cite{B59}), which motivated its
inclusion in our sample, but Bertaud \& Floquet (\cite{BF67}) classified it
A2-F2 (Am). Its classification remains ambiguous, and it would be interesting
to know its $\Delta a$ index in Maitzen's (\cite{M76}) photometry. Its
peculiarity index in Geneva photometry is $\Delta (V1-G) = -0.006$ only, which
is typical of normal stars, but the efficiency of this index is known to be low
for such cool Ap stars. This is an excentric SB2 system, whose inclination
angle $i$ remains unknown. We secured 60 points (Fig.~\ref{RVa}) and obtained
the orbital elements listed in Table~\ref{orbel}.

A rough estimate of the inclination angle $i$ and of the masses of the 
components can be done
using uvby$\beta$ photometry\footnote{$V=7.68$, $b-y=0.194$, $m_1=0.214$,
$c_1=0.679$, $\beta = 2.739$, by Cameron (\cite{C66})} and the calibration by
Moon \& Dworetsky (\cite{MD85}). The physical parameters obtained are listed in
Table~\ref{par22-56}. Combining these results with the models of
Schaller et al. (\cite{SS92}), one
finds an approximate mass ${\cal M} = 1.80 \pm 0.09 {\cal M}_\odot$ and an
inclination $i$ close to 75\degr.

\subsection{HD 65339 (= 53 Cam = BD +60\degr 1105 = Renson 17910)}

53~CAM is a very well studied A3 SrEuCr star (Osawa \cite{O65}). It is known as
a binary by both spectroscopy and speckle interferometry.
The speckle orbit was published by Hartkopf et al. (\cite{H96}) and a
radial-velocity curve was published by Scholz \& Lehmann (\cite{SL88}).
Combining our 46 measurements (Table~1) with those published by
Scholz \& Lehmann (\cite{SL88}), we determine the orbital
parameters listed in Table~\ref{orbel}. The scatter of the residuals of
Scholz's measurements
are similar to those of CORAVEL observations alone, which appears surprising
at first sight. Examining the depth and width of the correlation dip as
a function of the {\it rotational} phase ($P = 8.0267$~days), one clearly sees
a significant variation of both quantities (see Fig.~\ref{hd65}). The residuals
around the fitted RV curve also show a variation, which is related, therefore,
to the spots associated with a non-negligible $v\sin i$. 53 Cam is then a nice
example of an object displaying two variations simultaneously, one due to
rotation (with an amplitude of up to 7 \kms peak-to-peak) and the other due to
a binary companion.

\begin{table}
\caption{Speckle observations of HD 65339 used for the simultaneous fit
on the $RV$, speckle and parallax data. Since the errors are not
given by the authors, an error of 0.01 arcsec and of $2\degr$ has been
assumed on the separation and position angle respectively.}
\begin{center}
\begin{tabular}{rrrl}
Epoch  & $\rho$  &  $\theta$ & Source \\
(frac. year)&(arcsec)&($\degr$)& \\ \hline
1980.1561    &     0.044    &   336.4 &McAlister et al. \cite{McA83} \\
1984.0526    &     0.093    &   299.6 &McAlister et al. \cite{McA87} \\
1984.8463    &     0.091    &   307.3 &Balega$^2$ \cite{Ba87} \\
1985.1830    &     0.091    &   306.3 &Balega$^2$ \cite{Ba87} \\
1985.1858    &     0.088    &   308.7 &Balega$^2$ \cite{Ba87} \\
1986.7039    &     0.045    &   328.6 &Balega et al. \cite{Ba89} \\
1986.8894    &     0.0339   &   332.44&Hartkopf et al. \cite{H96} \\
1989.2267    &     0.063    &   283.1 &McAlister et al. \cite{McA90} \\
1990.2755    &     0.089    &   293.1 &Hartkopf et al. \cite{Hrt92} \\
1991.3265    &     0.086    &   303.8 &Hartkopf et al. \cite{Hrt94} \\
1991.8943    &     0.085    &   306.7 &Hartkopf et al. \cite{Hrt94} \\
1992.3124    &     0.080    &   310.7 &Hartkopf et al. \cite{Hrt94} \\ \hline
\end{tabular}
\end{center}
\label{speckle}
\end{table}
\begin{table*}
\caption{Orbital parameters of HD 65339 obtained with a simultaneous fit
on the $RV$, speckle and parallax data.}
\begin{center}
\begin{tabular}{|r|r|r|r|r|r|r|r|r|r|r|r|} \hline
\multicolumn{1}{|c|}{Star name} & \multicolumn{1}{c|}{$P$} &
 \multicolumn{1}{c|}{$T_\circ$ (HJD} & \multicolumn{1}{c|}{$e$} &
 \multicolumn{1}{c|}{$V_\circ$} &
 \multicolumn{1}{c|}{$\Omega_1$} &
 \multicolumn{1}{c|}{$\omega_1$} &
 \multicolumn{1}{c|}{$i$} &
 \multicolumn{1}{c|}{$K_{1,2}$} & \multicolumn{1}{c|}{${\cal
 M}_{1,2}$} & \multicolumn{1}{c|}{$a_{1,2}$} & \multicolumn{1}{c|}{$\pi$}
\\ &
 \multicolumn{1}{c|}{(days)} & \multicolumn{1}{c|}{$-2400000)$} & &
 \multicolumn{1}{c|}{($\mathrm{km\,s^{-1}}$)} &
 \multicolumn{1}{c|}{($^\circ$)} &
 \multicolumn{1}{c|}{($^\circ$)} &
 \multicolumn{1}{c|}{($^\circ$)} &
 \multicolumn{1}{c|}{($\mathrm{km\,s^{-1}}$)} & 
 \multicolumn{1}{c|}{(${\cal M}_\odot$)} &
 \multicolumn{1}{c|}{($10^{6}$~km)} &
 \multicolumn{1}{c|}{(mas)}\\ \hline
HD 65339  & 2418.9
 & 27738.8 & 0.742 & -1.94 & 116.80 & 7.30 & 134.3 & 12.33 & 1.49 & 275.1&10.2\\
\multicolumn{1}{|l|}{(RV}& 2.41 & 15.6 & 0.013 & 0.13 & 1.31 & 1.37 & 4.4 &
0.41 & 0.66 & &1.0\\
+speckle)&&&&&&&&&&&\\
          & 
 &         &       &       &        &      &       & 12.13 & 1.52 & 270.5&\\
            &      &      &       &      &      &      &     & 3.25 & 0.33 & &\\
 \hline
\end{tabular}
\end{center}
\label{orbel65}
\end{table*}

\begin{figure}
\resizebox{\hsize}{!}{\includegraphics[angle=-90,width=9.5cm]{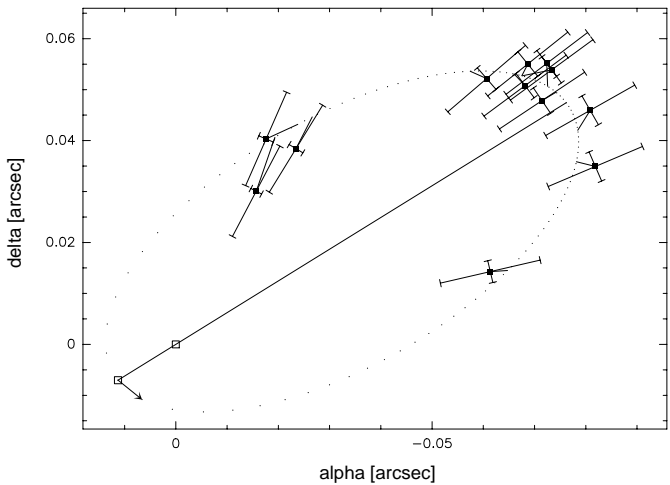}}
\caption{Orbit of the system HD 65339 projected on the sky, as fitted to both 
radial velocities and speckle measurements simultaneously. The axes are labeled
in arcseconds. The error bars have been chosen in an arbitrary way because
they are not given by the authors.}
\label{orb65}
\end{figure}

Thanks to a code made available by T. Forveille and developed in Grenoble,
we have fitted
{\it simultaneously} the radial velocities, the speckle measurements and the
Hipparcos parallax ($\pi = 10.16 \pm 0.77$), leaving not only $K_1$ but also $K_2$ as an adjustable
parameter in spite of the lack of $RV$ data for the companion. The speckle
measurements retained for the
fit are given in Table~\ref{speckle}, while the results are shown in
Table~\ref{orbel65}. The ``visual'' orbit is shown in Fig.~\ref{orb65}.

This is the first time that such a solution
is attempted for this system. The results are surprising, in that both
companions appear to have the same mass, contrary to what
Scholz \& Lehman (\cite{SL88}) had found (2.5~$\cal M_\odot$ and
1.6~$\cal M_\odot$ for the primary and the secondary respectively) by
combining the separate $RV$ and speckle orbits (they used a photometric mass
for the primary, since there was no good parallax value at the time). On the
other hand, they are compatible with the small $\Delta m$ required for speckle
observations. They also differ from the
mass estimate done by Martin \& Mignard (\cite{MM98}) on the basis of
Hipparcos results, which has a large error, however. The uncertainty is very
large and could be substantially reduced if the spectrum of the secondary
could be observed. We could not see it on our ELODIE spectra, but this is not
surprising since they were taken when both companions had almost the same
radial velocity.

\subsection{HD 73709 (= BD +20\degr 2165 = Renson 20510 = Praesepe KW 279)}

HD~73709 was classified A2-A5-F0 (Am) by Gray \& Garrison (\cite{GG89}), but was
found photometrically Ap by Maitzen \& Pavlovski (\cite{MP87}) according to the
$\Delta a$ index ($\Delta a = 0.018$). The Geneva peculiarity index gives an
ambiguous answer: $\Delta (V1-G) = 0.001$ is a few thousands of magnitude larger
than the average of normal stars, but is not conspicuous. It has been put lately
in our programme because of its photometric peculiarity, first for magnetic
field measurements, second for radial-velocity monitoring.

Two ELODIE data were taken in the course of the survey
for magnetic fields, while a third one has kindly been obtained for us by
Mr. Dominique Naef (Geneva Observatory) during a planet-search programme.

HD 73709 is extremely interesting because of its reliable Am classification
and positive $\Delta a$: it was generally accepted that Am stars never show
enhanced $\Delta a$ values (Maitzen \cite{M76}, Maitzen et al. \cite{MP98}) which are
characteristic of magnetic Ap stars only. Conversely, large-scale magnetic
fields are generally not found in Am stars, with the probable exception of
the hot Am star $o$ Peg (Mathys \cite{M88}, Mathys \& Lanz \cite{ML90}). The three spectra
taken with the ELODIE spectrograph consistently show a surface magnetic field
of about 7.5 kG which seems very significant, in spite of a relatively large
projected rotational velocity $v\sin i = 16$ \kms (Babel \& North, in
preparation).

The orbit of this star was published by Debernardi et al. (\cite{DM20}).
However, we have 12 additional data, so we have redetermined the orbit using
both published and new data (note that the data published by Debernardi et al.
have not been put into the ELODIE RV system (Udry et al. \cite{udry}), so that the RV values used here
are very slightly different from those published by these authors). The orbit
is slightly improved.

\subsection{HD 98088A (= BD -6\degr 3344A = Renson 28310)}
This is a well-known SB2 binary hosting a magnetic Ap star of the type SrCr
according to Osawa (\cite{O65}). Its binary nature has been discovered by
Abt (\cite{A53}),
who saw it only as an SB1, and the complete orbital solution of the SB2 system
was given by Abt et al. (\cite{A68}). These authors have shown that the spectral
variations have the same period as the orbital one and that the system must,
therefore, be synchronized. According to them, the spectral type of the primary
is A3Vp while that of the secondary is A8V. In spite of the binary nature of
this star, the Geneva photometric system ``sees'' its peculiarity, with
$\Delta (V1-G)=0.013$, and Maitzen's (\cite{M76}) photometry is even more efficient,
with $\Delta a=0.035$. Therefore, the primary is a rather extreme Ap star.

A very interesting feature of HD 98088 is that,
in spite of its relatively short orbital period, it has a significant
eccentricity, so that one may expect an apsidal motion to take place.
According to Wolff (\cite{W74}), only marginal evidence for such a motion could be 
found over a time base of 20 years, and new observations should be done 20 to
30 years later to settle the question, the expected period of the apsidal motion
being 500 to 700 years. Because of this expectation, we reobserved the system
with CORAVEL in the spring of 1998. Bad weather prevented us to obtain a dense
coverage of all phases, but 17 observations of the primary and 8 of the
secondary could be done. The period could be refined to
\begin{equation}
P = 5.905111 \pm 0.000004 ~{\mbox{days}}
\end{equation}
The $\omega$ angle has not changed in a significant way since about 30 years,
since we find $\omega_1=314.25\pm 3.66 \degr$, while the combined literature
data for the epochs 1953-1973 give $\omega_1=314.41\pm 1.09 \degr$ according to
Wolff (\cite{W74}). We verified this result with our code, which gives
practically the same value but a larger uncertainty
($\omega_1=314.47\pm 1.60 \degr$, for the primary RV curve alone).
If the period of the apsidal motion was 700 years as suggested by Wolff
(\cite{W74}), then the argument of the periastron should have changed by
$18\degr$ in 35 years, so we should have found $\omega_1\sim 332.5 \degr$.
This is five $\sigma$ away from our result, so we conclude that the apsidal
motion can only be much slower, with a period probably longer than a millenium.
If one imposes $\omega_1=332.5\degr$, the fit of the CORAVEL observations is
clearly worse, with an rms scatter of the residuals of 3.83\,\kms\ instead of
2.74 \kms\ (for both components); the difference is more visible on the RV
curve of the primary ($\sigma_{\mathrm{res}}=3.89$ \kms\ instead of $2.35$).
Combining all published data with the CORAVEL ones, and after a correction
$\Delta RV=-1.14$~\kms\, to the latter for a better consistency, we
obtain a very good curve with $\omega_1=314.46 \pm 1.44 \degr$ (Fig. \ref{RVb}).

Fortunately, this system has a rather good Hipparcos parallax of
$\pi = 7.75\pm 0.76$~mas, so that the radii of its components can be estimated.
From the observed apparent magnitude $V_{1+2}=6.107$ (Rufener \cite{Ru88}) and
from the magnitude difference $\Delta V=1.2$ (Abt et al. \cite{A68}) one gets
the individual apparent magnitudes $V_1=6.42$ and $V_2=7.62$ which give the
absolute magnitudes $M_{V1}=0.87$ and $M_{V2}=2.07$ using the Hipparcos
parallax. From the spectral types A3 and A8 proposed by Abt et al. (\cite{A68}),
a first guess of the effective temperatures is given by the calibration of
Hauck~(\cite{H94}): $T_{\mathrm{eff1}}=8275$~K and $T_{\mathrm{eff2}}=7532$~K.
Another guess can be done from the $(B2-G)$ index of Geneva photometry,
according to the calibration of Hauck \& North (\cite{HN93}): one has first to
subtract the typical Geneva colours of the companion (assuming an A8V star) to
the observed ones in order to get $(B2-G)_1=-0.455$, which corresponds to
$T_{\mathrm{eff1}}=8043$~K. Note that $(B2-G)_1$ is not very sensitive to the
assumption made on the companion, since it differs by only 0.023 mag from the
observed value $(B2-G)_{1+2}=-0.432$. Adopting this effective temperature for
the primary, an interpolation in the evolutionary tracks of Schaller et al.
(\cite{SS92}) for an overall solar metallicity yields the physical
parameters listed in Table~\ref{physHD98}. It is interesting to notice that the
mass ratio obtained in this way is $q=0.776\pm 0.049$, which is compatible to
better than one sigma with the dynamical mass ratio $q_{dyn}=0.737\pm 0.008$.

Also listed in Table~\ref{physHD98} are the radii estimated from the CORAVEL
projected rotational velocities assuming a negligible Zeeman broadening, from
the spin period (synchronization makes it equal to the orbital one) and from
$i=66\degr$.
The latter value is obtained from $M_1\sin^3i=1.733\pm 0.030$ with the mass of the primary interpolated in the
evolutionary tracks. It is almost
identical with $i=67\degr$ proposed by Abt et al. (\cite{A68}). The radii
obtained through the projected rotational velocities are compatible with those
obtained from the Hipparcos luminosity and photometric effective temperatures,
in the sense that error bars overlap. The agreement is perfect for the
secondary, but much less satisfactory for the primary, even though the
difference is less than twice the largest sigma.
\begin{table}
\caption{Physical parameters of both components of the spectroscopic system
HD~98088A, inferred from photometric temperatures and Hipparcos parallax.}
\begin{center}
\begin{tabular}{lcc}\hline
Parameter              & primary        & secondary \\ \hline
$M_V$                  & 0.87           & 2.07      \\
$\log(T_{\mathrm{eff}})$&$3.905\pm 0.016$&$3.877\pm 0.017$\\
$\log(L/L_\odot)$      &$1.60\pm 0.09$  &$1.10\pm 0.09$\\
$M$ ($M_\odot$)        &$2.261\pm 0.093$&$1.755\pm 0.085$\\
$R$ ($R_\odot$)        &$3.27\pm 0.43$  &$2.10\pm 0.28$\\
$\log g$ (cgs)         &$3.76\pm 0.10$  &$4.04\pm 0.11$\\ \hline
$R\sin i=\frac{P\cdot v\sin i}{50.6}$ ($R_\odot$)&$2.46\pm 0.25$&$1.84\pm
0.22$\\
$R=R\sin i/\sin(66\degr)$&$2.70\pm 0.27$  &$2.02\pm 0.24$ \\
$d$ (pc)               &\multicolumn{2}{c}{$129\pm 13$}\\ \hline
\end{tabular}
\end{center}
\label{physHD98}
\end{table}
An attempt has been made to impose the dynamical mass ratio $q_{dyn}=0.737$ and
interpolate in the evolutionary tracks the pair of stars whose magnitude
difference
is compatible with it. Maintaining the assumption of an A8V companion, we get in
this way $\Delta V=1.54$ and $M_1=2.30\pm .09$, $M_2=1.69\pm .08\,M_\odot$,
$R_1=3.42\pm 0.45$, $R_2=1.86\pm 0.25\,R_\odot$. The magnitude difference
appears a bit large compared with the estimate of Abt et al. (\cite{A68}) and
the radius of the primary turns out to be even larger, making the discrepancy
more severe compared to the radius estimated from the rotational velocity.

The number of CORAVEL measurements is too small to conclude about the possible
variability of the depth and width of the correlation dip of the primary.

\subsection{HD 105680 (= BD +23\degr 2423 = Renson 30570)}

This star was listed A3p SrSi? by Bertaud (\cite{B59}), which motivated its
inclusion in the sample, and as A3-F2 by Bertaud \& Floquet (\cite{BF67}). The
radial-velocity curve is very well defined (see Fig.~\ref{RVb}). We secured 42 points
over an interval of 2966 days. In spite of a relatively large
$v\sin i$, the rms scatter of the residuals is small. Unfortunately, the
classification remains ambiguous; $\Delta (V1-G) = 0.004$ suggests a mild
peculiarity, but it is not large enough to exclude that it may be an Am star
instead of an Ap.

\subsection{HD 138426 (= BD -18\degr 4088 = Renson 39420)}
This poorly known star has been classified Ap SrCr(Eu) by Houk \& Smith-Moore
(\cite{H88}). Its photometric peculiarity is just significant in the Geneva system
($\Delta (V1-G)= 0.010$) and it is clearly an SB1 binary with a relatively
short period. The $v\sin i$ is very small ($< 2.4$~\kms) and neither the depth 
nor the width of the correlation dip seems to vary. Figure~\ref{RVb} shows a
phase diagram of the radial velocities. The residual scatter is rather large,
but the most discrepant points (at phases 0.48 and 0.56) were observed in the
run of March 1997 where technical problems prevented the data to be registered
on tape, so that it has not been possible to evaluate their quality.

\subsection{HD 184471 (= BD +32\degr 3471 = Renson 50890)}

This star was classified A9 SrCrEu by Bertaud \& Floquet (\cite{BF74}). A total of 36
measurements have been made over almost 3500 days (Table~1),
which clearly define a 429-day period (see Fig.~\ref{RVb}). The residuals are very
small thanks to a small $v\sin i$ ($< 2$ \kms) and a well contrasted dip.

\subsection{HD 188854 (= BD +46\degr 2807 = Renson 52220)}

Ap or Am, according to different authors, its spectral type is not well
determined. HD~188854 was listed as A7 CrEu by Bertaud \& Floquet (\cite{BF74}), but
also as A5-F0 (Bertaud \& Floquet \cite{BF67}). No $\Delta a$ photometry has been
published for it, and the Geneva index $\Delta (V1-G) = -0.002$ does not allow us
to conclude, especially as it is among the coolest existing Ap stars.
The radial-velocity curve is well determined with a $\sigma(O-C)$ of $0.51$~\kms\
only (see Fig.~\ref{RVb}).

\subsection{HD 200405 (= BD +47\degr 3256 = Renson 55830)}
This A2 SrCr (Osawa \cite{O65}) star had already been announced as having the shortest
orbital period known among all Bp and Ap stars (North \cite{N94}), with a period of
only 1.635 days. A survey of the literature has not denied this claim: the few
binaries with a period shorter than 3 days in Renson's (\cite{R91}) catalogue either
owe their spectral peculiarity to another physical cause like {\it in situ}
nucleosynthesis (HD 93030, an "OBN" star according to
Sch\"onberner et al. \cite{SH88}
and HD 49798, an O6 He star), or are misclassified (HD 25833, a normal B4V star
according to Gimenez \& Clausen \cite{GC94}), or do not have a typical Bp, Ap
peculiarity (HD 124425, F7 MgCaSr in Renson's catalogue; HD 159876, F0IIIp in
the Hipparcos Input Catalogue but A5-F1 $\delta$ Del? in Renson's, and
Am, A7/A9/F3 according to Abt \& Morrell \cite{AM95}); finally, the A2 CrEu star
HD 215661B is not a binary: only the A component of this visual system is an
Algol-type binary.

HD 200405 is a {\it bona fide} Ap star also from the photometric point of view:
Geneva photometry shows it is peculiar, with $\Delta(V1-G)=0.021$, and Maitzen's
peculiarity index $\Delta a=0.038$ on average (Schnell \& Maitzen \cite{SM95}).

The inclination angle $i$ of the orbital plane of HD 200405 must be very small,
according to the value of $a_{1} \sin i$ and of the mass function
(Table~\ref{orbel}), unless the companion is a brown dwarf.
The radial-velocity curve is shown in Fig.~\ref{RVb}. The orbit is circular ($e = 0$).
This object is especially interesting, since it is exceptional: all other
binaries with a magnetic Ap component have orbital periods longer than 3 days.
If tidal effects tend to wash out the chemical peculiarity of the components,
as suggested by this lower limit, then one has to explain how HD~200405 has been
able to remain an Ap star in spite of significantly large tides.

Another way to interpret this radial-velocity curve would be to assume that 
HD 200405 has a very small, highly contrasted spot with enhanced abundance of
iron-peak elements (whose lines are selected by the CORAVEL mask); in such a
case, rotation alone might be responsible for a sinusoidal curve, if both the
inclination $i$ of the rotational axis and the angle between the rotation and
spot axes are such that the spot remains visible during the whole cycle.
However, such a situation appears extremely improbable, since one does not see
any variation in the intensity of the correlation dip, nor in its width or
depth, which should occur because of the varying aspect of the spot. Likewise,
the radial velocity of the H$_\alpha$ line measured once with ELODIE is
compatible with the CORAVEL RV curve, while one would rather expect it to
remain at the "systemic" velocity. Furthermore, the spot hypothesis would imply
an {\it apparent} $v\sin i \approx 0$~\kms (the radial velocity of every point
in the spot being practically the same), while we obtain
$v\sin i = 9.5\pm 0.4$ and
$7.9\pm 0.2$ \kms using respectively CORAVEL and ELODIE: a small spot could
never give rise to such a high value (the effect of the magnetic field has been
removed in the ELODIE estimate). Therefore, HD 200405 holds the record of the
shortest orbital period known.

\subsection{HD 216533 (= BD +58\degr 2497 = Renson 59810)}
This A1 SrCr star (Osawa \cite{O65}) was already known as an SB1 system.
Floquet (\cite{F79}) found an orbital period of 16.03 days, using the radial velocity
of the Ca\,{\sc ii} K line.

A total of 48 measurements have been made over almost 6225 days. The
16.03-day period does not fit at all our radial velocities. We find a much 
longer period $P = 1413$ days (see Figure~\ref{RVb}), which should be considered as 
more reliable. It seems that Floquet was too confident in her assumption of an
homogeneous distribution of ionized calcium on the surface of the star, and that
the RV variation she observed was in fact due to a spot. The rotational
period of this star, 17.2 days, is indeed very close to the ``orbital'' one
found by Floquet (\cite{F79}), although not identical.

\section{Ap stars with incomplete orbital RV curve or with possible intrinsic
variability}

Most stars presented in this section have a $P(\chi^2)< 0.01$, so they may
be binaries; others, with $P(\chi^2)> 0.01$ are also discussed because they are
suspected binaries from the literature. Although some of these are obviously
double stars, others surely
owe their variability to the effect of spots and relatively rapid rotation.
These stars should remain under monitoring, not so much to confirm the cause of
their variability, which seems clear, but rather to look for any long term
variation which would betray the existence of a companion. Although identifying
spotted stars may appear problematic at first sight, one sees {\it a posteriori}
that in general CORAVEL yields enough data to disentangle them:
first, they always have a significant $v\sin i$, generally above 15 \kms;
second, their range of variability is relatively small, i.e. typically 10\,\kms\
and at most 20 \kms; third, the timescale of the variation is small and
compatible with the rotational period expected from $v\sin i$ and the oblique
rotator formula $v\sin i = 50.6 R\sin i/P_{rot}$. Another way to summarize these
criteria is simply to look at the mass function that would result from
interpreting the variations as due to a companion: it is always extremely small.
In addition, a significant variation of the depth and width of the correlation
dip betrays the existence of spots; if these values remain constant, then any
RV variation appears more probably due to a companion.
In some cases of course, ambiguity will remain. On the other hand, the choice
we made of not rejecting spotted stars {\it a priori} (contrary to what
e.g. Abt \& Snowden \cite{AS73} did) has the advantage of drawing the attention to 
potentially interesting objects for future Doppler-imaging investigations.

In the following, Ap stars which present a RV variation but for which no
orbital solution could be found are presented. Some simply have too long an
orbital period for even one cycle to be covered. Others are probably spotted
stars and are discussed by examining not only their RV variation, but also
the variation of the width and depth of their CORAVEL correlation dip. A few
stars with fast rotation and only 2 to 5 measurements are not discussed in
spite of their very small $P(\chi^2)$: their correlation dips are so shallow
that the internal errors are probably underestimated, so their variability
remains uncertain. HD 157740, 216931 and 220825 are in this case.

\subsection{HD 2453 (= BD +31\degr 59 = Renson 560)}
\begin{figure*}
\centering
 \includegraphics[width=17cm]{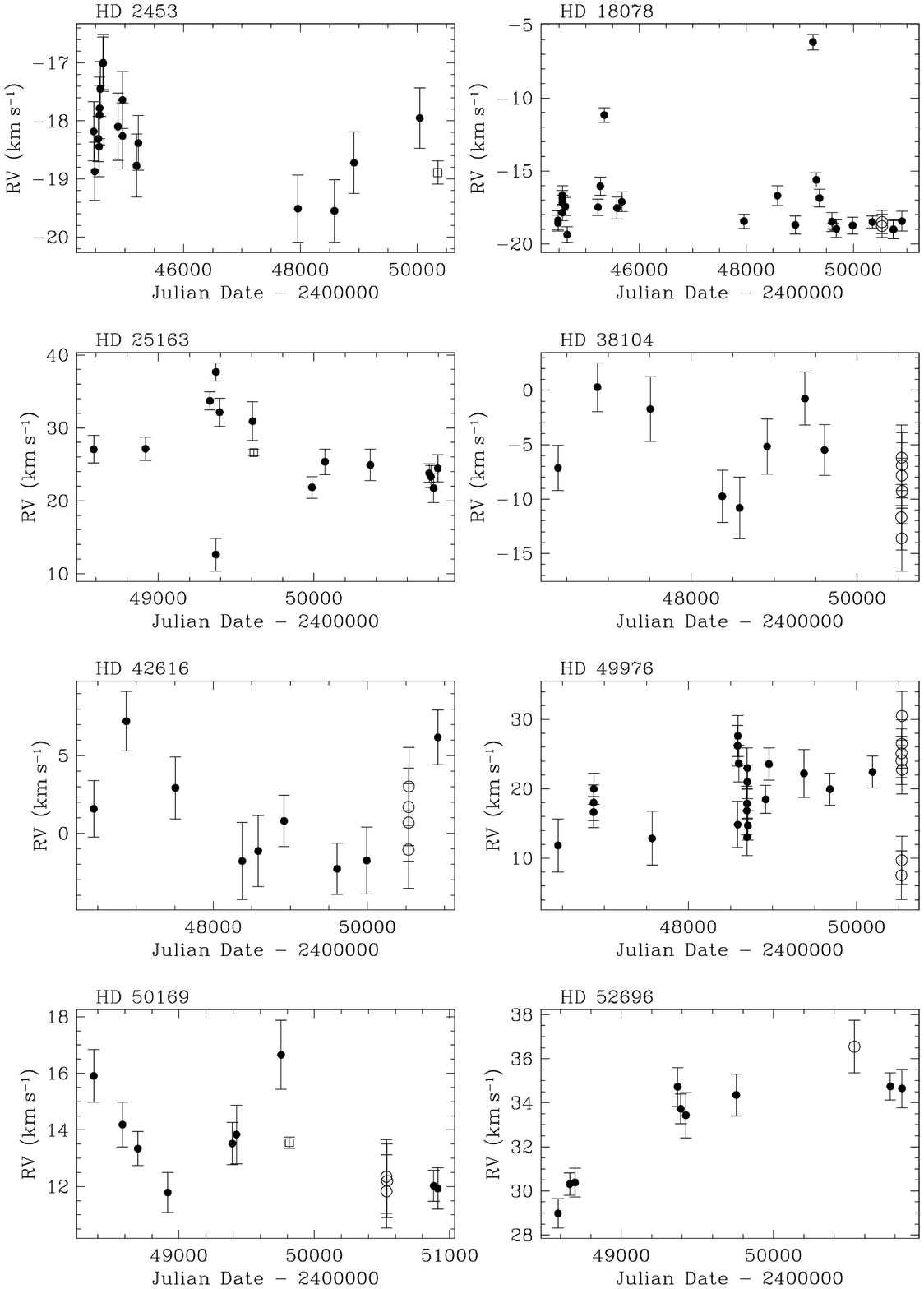}
 \caption{Radial velocity vs HJD for all stars which seem variable but for which
 no orbital solution could be obtained. Many of these objects probably owe their
 variability to intrinsic spectroscopic variations (rotation and spots).
 Full dots are CORAVEL data, while open squares are ELODIE data; open dots
 represent CORAVEL data registered by hand in the dome during an observing run
 (22 March to 2 April 1997) when the data were not properly printed on tape.}
 \label{RVc}
\end{figure*}

This object has been classified Ap SrCrEu by Osawa (\cite{O65}). It is a rather
extreme Ap star, with $\Delta (V1-G) = 0.038$.
The radial velocities are represented as a function of time in Figure~\ref{RVc}. There
are significant variations, but there are not enough data to determine the period.

\subsection{HD 18078 (= BD +55\degr 726 = Renson 4500)}
This star was classified Ap SrCrEu by Osawa (\cite{O65}) and was listed as SrSi? in
the compilation of Bertaud (\cite{B59}). Geneva photometry shows it is an extreme
Ap star, with $\Delta (V1-G) = 0.047$.
Figure~\ref{RVc} shows the radial velocities versus time. Two points are much higher
than the others, suggesting a very excentric orbit with a possible period of
about 978 days. Interestingly, the Hipparcos parallax of this star is
$\pi = 0.51\pm 1.00$~mas, which is surprisingly small: this 8.3 magnitude star
is expected to lie at no more than about 300 pc. One might think that the
unknown duplicity has biased the astrometric solution, but the latter seems
satisfactory since no G or X flag appears in the Hipparcos Catalogue. On the
other hand, the parallax listed is only about $2.8\,\sigma$ from the expected
one, a non-negligible possibility.
\begin{figure*}
\centering
 \includegraphics[width=17cm]{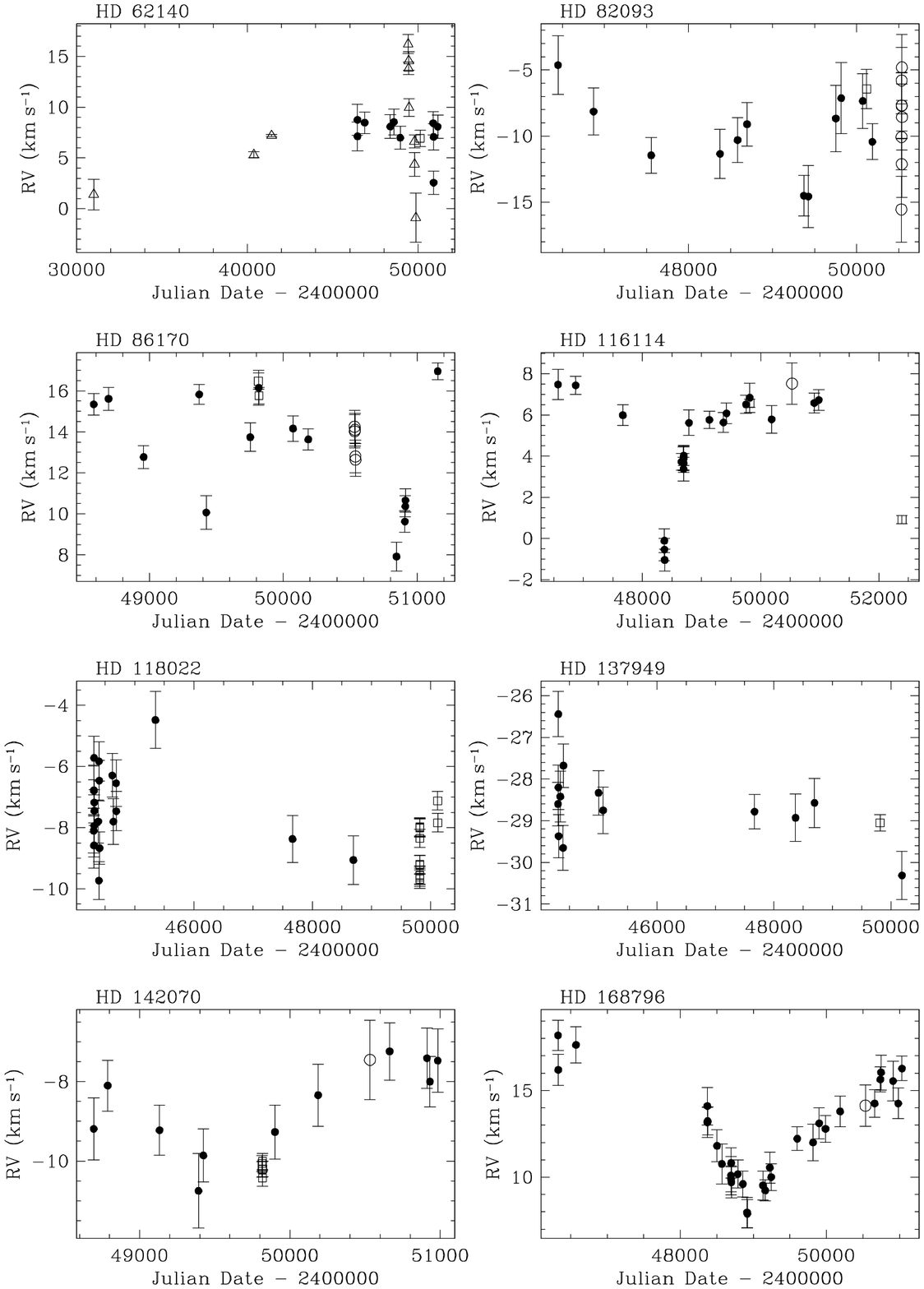}
 \caption{Same as Fig. 11. Open triangles represent literature data. For
 HD~62140 (49 Cam), the triangles are the data considered by Mkrtichian et al.
 \cite{MS97}. Notice that HD~116114 and HD~168796 are clearly
 long-period binaries with an eccentric orbit.}
 \label{RVd}
\end{figure*}

\subsection{HD 25163 (= BD -4\degr 709 = Renson 6400)}
This star, classified Ap SrCr by Bidelman \& McConnell (\cite{BM73}), does not appear
conspicuous in Geneva photometry, since its peculiarity parameter is
$\Delta (V1-G) = -0.006$, the same as for normal stars. It might be a
spectroscopic binary, since its radial velocity is variable as shown in
Figure~\ref{RVc}, and a marginal secondary dip has been observed once. On the other hand,
$v\sin i \leq 20.9\pm 2.1$ suggests that spots might be the cause of the variation.
Since the secondary dip was observed only once and is very shallow, we tend to
dismiss it as not significant, and to favour the spot hypothesis. This is an
ambiguous case which merits additional data.

\subsection{HD 38104 (= BD +49\degr 1398 = Renson 10240)}
This star was classified Ap Cr by both Osawa (\cite{O65}) and Cowley et al.
(\cite{CC69}).
It has a moderate photometric peculiarity, with $\Delta (V1-G) = 0.009$, which
is marginally significant. Its projected rotational velocity is relatively fast,
$v\sin i \leq 24.4\pm 2.4$, and the variability timescale is quite short
(Fig.~\ref{RVc}).
Therefore, spots may cause the RV variation, especially as both
the depth and the width of the correlation dip seem variable too.

\subsection{HD 42616 (= BD +41\degr 1392 = Renson 11390)}
This star was classified CrSrEu by Osawa (\cite{O65}) and has a significant
photometric peculiarity ($\Delta (V1-G) = 0.022$). It is a photometric
variable, but the only lightcurves published so far are those of Rakosch \&
Fiedler (\cite{RF78}); their 17.0~days period is not in agreement with
our radial velocity data. The small $RV$ variations shown in
Figure~\ref{RVc} are probably due to spots and rotation,
though an orbital motion cannot be excluded without additional data.

\subsection{HD 47103 (= BD +20\degr 1508 = Renson 12630)}
The strong magnetic field of this SrEu star was discovered by Babel et al.
(\cite{BNQ95}) with a correlation technique applied to ELODIE spectra, and
Babel \& North (\cite{BN97}) noticed a probable slight $RV$ shift from one year
to another. The unique CORAVEL measurement seems to confirm this variability,
though additional data would be needed to confirm it.

\subsection{HD 49976 (= BD -7\degr 1592 = Renson 13560)}
Osawa (\cite{O65}) classified this star Ap SrCr. It is photometrically peculiar too,
with $\Delta (V1-G) = 0.025$ and $\Delta a = 0.045$ mag. Its rotational period
is $P_{\mbox{rot}} = 2.976$~days (Pilachowski et al. \cite{PB74},
Catalano \& Leone \cite{CL94}).
There are significant RV variations on a short timescale,
as Figure~\ref{RVc} shows, which are probably
due to spots and rotation, especially as it is a known spectroscopic variable
(Pilachowski et al. \cite{PB74}): the depth of the dip is variable, while its width
does not seem to vary in a significant way.

\subsection{HD 50169 (= BD -1\degr 1414 = Renson 13700)}
This is a classical Ap SrCr star (Osawa (\cite{O65}) with a 5~kG surface
magnetic field (Mathys et al. \cite{MHL97}). The latter authors suspected it of
being a long period binary, saying ``Between our first and last observations of
HD 50169 (4 years apart), its radial velocity has monotonically increased, by
about 2~\kms''. Although our RV values (shown in Figure~\ref{RVc})
indeed suggest that it is a binary, they are
difficult to reconcile with this statement: they suggest a period of roughly 4
years. The mean of 3 $RV$ measurements made by Grenier et al. (\cite{G99}) is
$+6.6\pm 1.0$~\kms, which seems significantly different from our values.
Unfortunately, these authors have not published the individual measurements.
On the other hand, 3 individual $RV$ values are listed by
Fehrenbach et al. (\cite{F97}): they range from 34~\kms\ on JD 2436525.602 to 23
and 25~\kms\ on JD 2436526.603 and JD 2438430.405 respectively. Even though the
accuracy of these objective-prism determinations is poor (4 \kms\ for an average
of 4 measurements), their average value is about $3\,\sigma$ higher than ours.
In any case, the $RV$ variation cannot be caused by spots because of the very
long rotational period (much longer than 4 years according to
Mathys et al. \cite{MHL97}, which is consistent with the lack of photometric
variability reported by Adelman et al. \cite{A98}).

\subsection{HD 52696 (= BD -19\degr 1651 = Renson 14460)}
Bidelman \& McConnell (\cite{BM73}) have classified this star as Ap SrEu. Geneva
photometry shows a slight, marginally significant $\Delta (V1-G) = 0.008$.
Maitzen's $\Delta a = 0.023$ mag (Maitzen \& Vogt \cite{MV83}) shows a better 
sensitivity than $\Delta (V1-G)$ for cool Ap stars.

The radial velocities are shown in Figure~\ref{RVc} as a function of time. HD 52696
is a slow rotator ($v\sin i \la 3.6\pm 1.9$ \kms\ according to CORAVEL) and the
correlation dip is quite well defined. There is no significant variation of
either the depth or the width of the dip. A steady increase of the radial 
velocity is evident and betrays the presence of a companion on a long-period 
orbit.

\subsection{HD 62140 (= 49 Cam = BD +63\degr 733 = Renson 17050)}
This classical Ap SrEu star (Cowley et al. \cite{CC69}) has a moderate $v\sin i$ of
about 20 km\,s$^{-1}$ and our CORAVEL data are entirely consistent with our
single ELODIE radial velocity. There is no indication of any $RV$
variation nor of any change of depth or width of the correlation dip.

However, Mkrtichian et al. (\cite{MS97}, \cite{MS99}) have presented
observational evidence for
a long-term $RV$ variation, which seems to betray the binary nature
of this star. Since the variation occurs on a time which is short compared
with the time of constant $RV$, the eccentricity of the orbit must
be high, if this is indeed a binary. Figure \ref{RVd} shows $RV$ as a function
of time for our data as well as for those of Mkrtichian et al. (\cite{MS97}),
who also include older data from the literature. If Mkrtichian's data are
right, then the eccentricity must really be very close to one in order to
explain such a sudden $RV$ excursion. More data taken at the critical epoch
would be interesting, if they exist at all.

\subsection{HD 81009A (= HR 3724A = BD -9\degr 2816A = Renson 22990)}
This Ap SrCrEu star (Cowley et al. \cite{CC69}) is a close visual and speckle binary
with a typical separation of 0.1 arcsec and a magnitude difference of about
0.1 (Renson \cite{R91}). This explains the rather marginal photometric peculiarity
of this star ($\Delta (V1-G)=0.006$), since its energy distribution is mixed 
with that of its normal companion. The orbital period of this system was not
well known at the time of the early observations, and since the radial velocity
appeared to remain constant
between 1980 and 1993, we thought it useless to observe it again. This was
a pity, since interesting things precisely began to happen in the last years;
thanks to Dr. Gregg Wade, who reminded us of this system, we observed it again
in spring 1998 and found that $RV$ had dropped from about 28 to 15
km\,s$^{-1}$. We certainly see the Ap component with CORAVEL, the other A star
having too few -- and probably too broad -- lines. Therefore, the system has
recently passed through a periastron, and additional data taken with other
instruments were used together with the CORAVEL ones by Wade et al.
(\cite{WD00}) to constrain the orbit. These authors found an orbital period of
29.3 years and also discuss the rotational period (33.984 days) and magnetic
geometry of the Ap component.

\subsection{HD 82093 (= BD -16\degr 2804 = Renson 23340)}
Bidelman \& McConnell (\cite{BM73}) classified this poorly known star as Ap SrCrEu (F).
It is clearly peculiar according to Geneva photometry, with
$\Delta (V1-G)=0.024$, and also according to Maitzen's photometry, with
$\Delta a = 0.041$ (Maitzen \& Vogt \cite{MV83}). In view of the large projected
rotational velocity ($v\sin i = 21.4\pm 2.1$~\kms), short timescale and small
amplitude of the $RV$ variation (Fig.~\ref{RVd}), it seems that it cannot be considered as
a binary and that spots are causing the variability. The depth of the 
correlation dip is definitely variable, but its width is only marginally so.
A period search on the $RV$ data yields no
clear-cut result, since many periods are possible. This is not so surprising,
since a $RV$ curve due to spots would be strongly anharmonic.
A similar period search has been attempted using the depth of the correlation
dip, but here again the results are ambiguous though they do indicate timescales
of a few days.

\subsection{HD 86170 (= BD -1\degr 2324 = Renson 24620)}
Bidelman (\cite{B81}) has classified this poorly known star as Ap SrCrEu. Its
photometric peculiarity is hardly significant in the Geneva system:
$\Delta (V1-G)=0.002$; unfortunately, there is no $\Delta a$ photometry
of it. It yields a very sharp correlation dip, so 
its projected rotational velocity is very small: $v\sin i\la 2$~\kms.
On the other hand, there are significant RV variations with a small
amplitude (Fig.~\ref{RVd}), which do not seem related with an effect of spots because
of the sharpness of the dip, unless the spot is very small and contrasted.
The depth of the correlation dip is definitely
not variable, while its width might be marginally so. No satisfactory period
could be found on the basis of the 21 available data.

\subsection{HD 116114 (= BD -17\degr 3829 = Renson 33530)}
Bidelman \& McConnell (\cite{BM73}) classified this star as Sr (F), while Houk
\& Smith-Moore (\cite{H88}) classified it Ap Sr(EuCr). Since it is very cool,
its photometric peculiarity is not large in the Geneva system:
$\Delta (V1-G)=-0.002$. On the other hand, Maitzen's photometry gives
$\Delta a = 0.011$, which is just significant.

As seen in Figure~\ref{RVd}, there is clearly a long-term variation of the radial 
velocities, which is typical of an eccentric binary.
The orbital period seems to be about
4000 days, which is very interesting because it contradicts the hypothesis
made in the Hipparcos catalogue (ESA \cite{esa}) of a 1014-day period.
Such a hypothesis can be understood because the passage at the periastron
occured at about JD 2\,448\,400, right in the middle of the Hipparcos mission.
The simple (but extremely improbable, even for a period of 1014 days!)
hypothesis of a null eccentricity then naturally leads to an underestimate of
the period.

There seems to be no conspicuous spot on this star. The apparent
$v\sin i=8.5\pm 0.3$~\kms appears significant but moderate, and possibly
due entirely to the Zeeman effect. The depth of the dip
might vary slightly, but its width remained constant.

\subsection{HD 118022 (= BD +4\degr 2764 = Renson 34020)}
This star, classified Ap SrCr by Osawa (\cite{O65}), has been the first one where
a magnetic field has been detected (Babcock \cite{B47}). Its photometric peculiarity
is strong in both Geneva ($\Delta (V1-G)= 0.029$) and Maitzen's (\cite{M76}) systems
($\Delta a = 0.048$). It does not seem to be a
binary, in spite of significant RV variations (Figure \ref{RVd}): both the width and depth of the correlation dip are strongly variable and the apparent projected
rotational velocity is significant. This is
a clear example of the effects of spots and rotation.

\subsection{HD 137949 (= BD -16\degr 4093 = Renson 39240)}
This star, classified Ap SrCrEu by Osawa (\cite{O65}), is rapidly oscillating with
a period of 8.3 minutes (Kurtz \cite{K82}, \cite{K91}). The photometric peculiarity is
zero in the Geneva system ($\Delta (V1-G)=-0.006$) but significant in
Maitzen's system ($\Delta a = 0.025$, Maitzen \cite{M76}; $\Delta a = 0.020$,
Maitzen \& Vogt \cite{MV83}). It shows small but significant RV variations
(Figure \ref{RVd}) which seem to be due to spots, since they occur on a rather
short timescale and the apparent $v\sin i$ is not negligible; furthermore,
there is a significant width variation
of the dip, although its depth does not vary.

\subsection{HD 142070 (= BD -00\degr 3026 = Renson 40330)}
This Ap SrCrEu star (Bidelman \& McConnell \cite{BM73}) has an average surface
magnetic field of 4.9 kG according to Mathys et al. (\cite{MHL97}), and was announced
as a spectroscopic binary by these authors. They also found a well defined,
relatively short rotational period $P_{\mbox{rot}}=3.3748\pm 0.0012$~days,
indicating a very small inclination of the rotation axis $i\la 8\degr$
because of the small $v\sin i$ implied by the sharpness of the lines
($< 5$~km\,s$^{-1}$). They observed a $RV$ variation of only
2.5~km\,s$^{-1}$ in about 500 days and concluded that the orbital period must
be, therefore, longer than this value. Indeed, our data, plotted in Fig.~\ref{RVd},
point to a period of about 2500 days or longer. Since a whole cycle has
probably not been completed yet, the uncertainty on the period remains too
large to give reliable orbital elements.

\subsection{HD 168796 (= BD +13\degr 3612 = Renson 47310)}
Classified Ap SrCrEu by Osawa (\cite{O65}), this poorly known star has a very
significant photometric peculiarity in the Geneva system
($\Delta (V1-G)= 0.022$) but has no $\Delta a$ value. Although its
apparent $v\sin i$
is far from negligible, there is no significant sign
of spots although the depth of the correlation dip might be slightly variable.
Figure \ref{RVd} shows that it is an eccentric binary
with a period longer than 5000 days.

\subsection{HD 171586A (= BD +4\degr 3801 = Renson 48090)}
Osawa (\cite{O65}) classified this star Ap SrCr. This peculiarity is confirmed
photometrically in the Geneva system ($\Delta (V1-G)=0.016$) and in Maitzen's
(\cite{M76}) system ($\Delta a=0.034$). The few CORAVEL observations,
especially the early ones, suggest a marginal variability probably caused by
spots and rotation (see Fig.~\ref{RVe}).

\subsection{HD 180058 (= BD -11\degr 4921 = Renson 50000)}
Bidelman \& McConnel (\cite{BM73}) have classified this star Sr (F). Geneva
photometry gives an ambiguous diagnostic of peculiarity ($\Delta
(V1-G)=0.003$); there is no measurement in Maitzen's system.
The few CORAVEL observations shown in Figure~\ref{RVe} would be compatible with a constant $RV$, except
for one discrepant point. Spots may be responsible for the variability.

\subsection{HD 190145 (= BD +67\degr 1216 = Renson 52790)}
This is an ambiguous case, listed SrSi ? in the compilation of
Bertaud \& Floquet (\cite{BF74}), but also Am. Since $\Delta (V1-G)= -0.006$, it
seems that the Am classification is more probable, although this test is not
very compelling. Unfortunately, there is no published $\Delta a$ value of this
star. There seems to be some RV variability, since $P(\chi^2)=0.004$, but 
most of it is due to the first two measurements made in September 1985 and more 
data are obviously needed (see Figure \ref{RVe}). If it is an Am star, and if the 
apparent variability is not due to a binary component, then this would be one
more single Am star, contradicting the idea that all Am stars are members of
binaries. In that case, however, intrinsic variability could only be due to
$\delta$ Scuti type pulsations, which occur on timescale of hours, 

The apparent projected rotational velocity is moderate and the
profile of the correlation dip does not vary, making the existence of abundance
patches unlikely. This would fit well the Am classification.

\subsection{HD 191654 (= BD +15\degr 4071 = Renson 53420)}
\begin{figure*}
\centering
 \includegraphics[width=17cm]{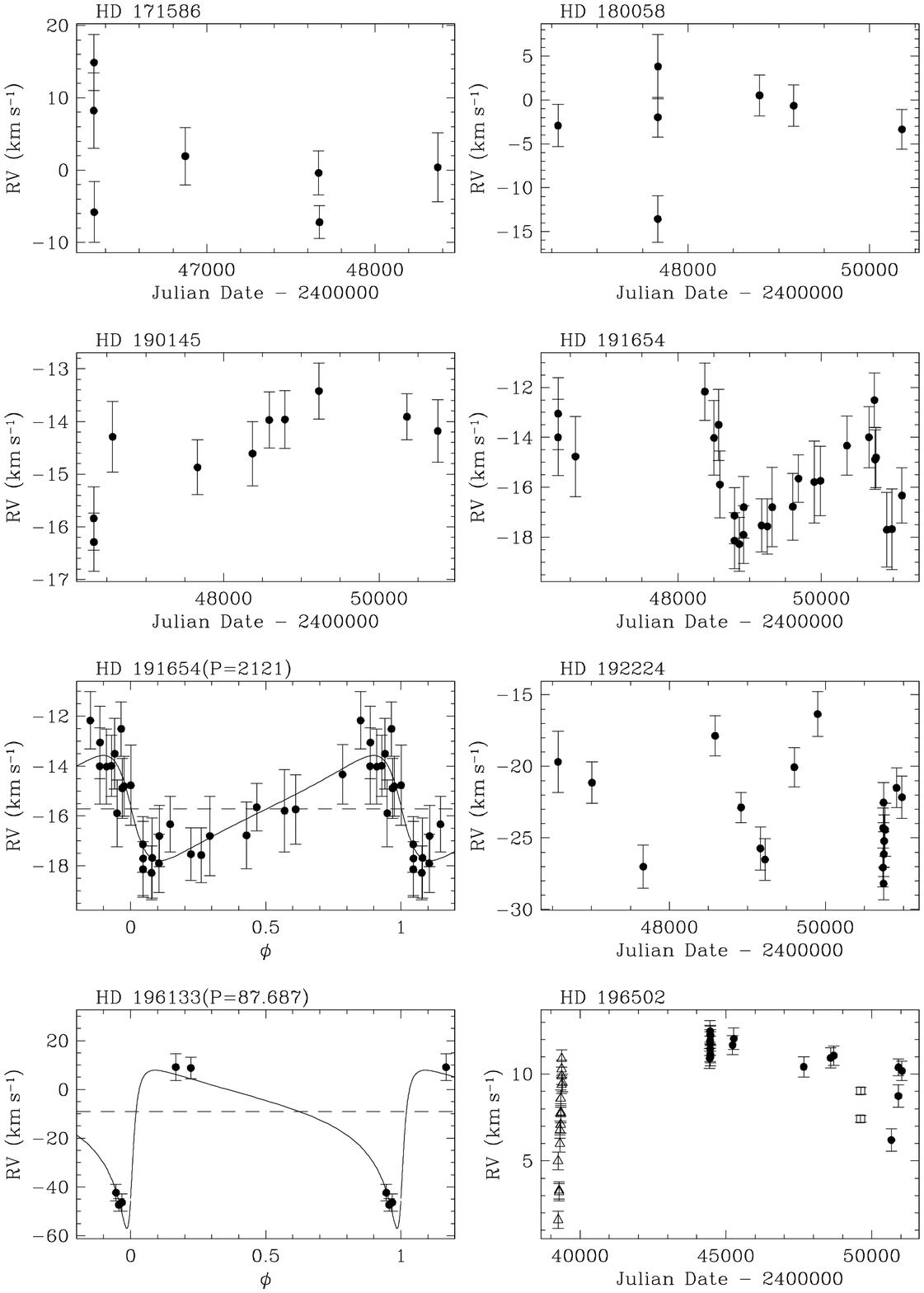}
 \caption{Same as Fig. 11. HD 191654 could be a SB1 with a long
 period $P=2121$ days. The bottom panel shows data
 from the literature as well as from CORAVEL and ELODIE, for the star HD 196502
 (same keys to the symbols as in Fig. 11).}
 \label{RVe}
\end{figure*}

Osawa (\cite{O65}) classified this star Ap SrCr and Geneva photometry nicely confirms
its peculiarity with $\Delta (V1-G)= 0.020$, though with a large uncertainty
(Rufener's \cite{Ru88} photometric weight $P$ is only 1). It is an interesting case
because it seems to rotate fairly fast while
showing, at first sight, no sign of short-term RV variations expected from
the effect of abundance
patches. The depth of the correlation dip varies slightly, but not its width.
On the other hand, a long-term RV variation seems to be present, which, if
confirmed, would betray the presence of a faint companion on an eccentric orbit
(see Fig.~\ref{RVe}). 
However, a short period very close
to one day is also possible, which would then correspond to
rotation and spots, not to any binary motion. A period search with Renson's
(\cite{R78}) method yields a possible period at $P_{\mbox{rot}}=0.99778$~days,
which leads to a
reasonable phase diagram. Such a short period would imply a
fast equatorial velocity and the star should have a nearly pole-on aspect
(an improbable occurrence) to explain the low $RV$ amplitude, if the
spot explanation were maintained.

The only existing estimate of the rotation period of this star has been
published by Vet\"{o} et al. (\cite{V80}): they found
$P_{\mbox{rot}}=1.857\pm 0.010$~days, on the basis of 13 differential photometric 
$u$ measurements. We have repeated the period search using their published
data (after correcting what appears to be a typo error at JD 2444092.373:
the magnitude difference is $-0.762$, not $-0.682$) and obtain
$P_{\mbox{rot}}=1.853$~days with Renson's method, which entirely confirms their
result. The radial velocities phased according to the photometric period give
only a scatter diagram, even though the photometric period is not far from twice
the possible short $RV$ period. Thus the binary hypothesis remains
possible (the corresponding orbital elements are listed in Table~\ref{hd191}), though it
is intriguing that the standard deviation of the $O-C$ residuals ($0.91$~\kms)
is smaller than the average error on individual data ($\sim 1.3$~\kms).
Additional measurements would be useful to conclude.
\subsection{HD 192224 (= BD +40\degr 4059 = Renson 53580)}
This star, classified Ap CrEu by Bertaud (\cite{B65}) and photometrically peculiar
according to the Geneva system ($\Delta (V1-G)= 0.028$), seems to satisfy all
the criteria of a spotted rotator: $v\sin i\leq 22.12\pm 2.21$~\kms, a $RV$
variability with a small 10~\kms~ peak-to-peak amplitude on a timescale of a
few days, and a significant variation of the depth of the correlation dip.
It has not been possible to find a convincing orbital solution, so it appears
probable that this is indeed a single star with abundance patches. The RV data
are shown in Fig.~\ref{RVe}.

\subsection{HD 196133 (= BD +44\degr 3505 = Renson 54660)}
The peculiarity of this star, classified Ap SiSr by Musielok (\cite{Mu76}), is
considered doubtful by Renson (\cite{R91}). Its Geneva peculiarity index
$\Delta (V1-G)=0.000$ does not help to solve the question, and this star was
not measured in Maitzen's photometric system. This is a known spectroscopic
binary with $P_{\rm orb}=87.687$~d,
whose elements have been published by Northcott (\cite{N48}). The CORAVEL data
are compared with this published orbit in Fig.~\ref{RVe}: the agreement is very
good, so that even the epoch of periastron passage does not need any
adjustment.

\subsection{HD 196502 (= 73 Dra = BD +74\degr 872 = Renson 54780)}
This star, classified Ap SrCrEu by both Osawa (\cite{O65}) and
Cowley et al. (\cite{CC69}),
is a ``classical'' one; it has a significant photometric peculiarity
($\Delta (V1-G)= 0.022$, $\Delta a = 0.072$: Maitzen \& Seggewiss \cite{MS80}).
The 18 CORAVEL data obtained to date show a constant radial velocity, except for
the measurement made in August 1997 (JD 2450668.46, see Fig.~\ref{RVe}). This confirms
the suspicion of Preston (\cite{P67}) who found it slightly variable on long
timescale. Preston
proposed a period of 551 days, but it does not fit well our data. Although
$v\sin i\leq 8.3\pm 0.3$~\kms and both the depth and the width of the
correlation dip are variable, spots can probably not explain the variability
because they would imply a short timescale which is definitely not observed.
A companion on a very eccentric orbit with a long period is probably needed
to explain the observed variations, part of which (e.g. the difference between
the two ELODIE results) might be due to the effect of spots and rotation.

\subsection{HD 201601 (= BD +9\degr 4732 = Renson 56210)}
This classical Ap star was classified SrCrEu by Osawa (\cite{O65}) and has a
significant photometric peculiarity, but only in Maitzen's (\cite{M76}) system:
$\Delta a= 0.011$. Interestingly, we find it perfectly constant while Scholz et
al. (\cite{S97}) found $RV$ to be about 10 km\,s$^{-1}$ higher than
previously from JD 2450353 to JD 2450356. There must have been some problem with
the data of Scholz et al., since measurements made by Mkrtichian and coworkers,
made with an instrument similar to CORAVEL, confirm our results. The whole
question raised by Scholz et al. (\cite{S97}) is discussed in more details by
Mkrtichian et al. (\cite{MS98}).

\subsection{HD 206088 (= BD -17\degr 6340 = Renson 57390)}
This star was classified Ap Sr by Bertaud \& Floquet (\cite{BF74}), but Am by
Osawa (\cite{O65}). It is not peculiar in the Geneva photometric system
($\Delta (V1-G)= -0.003$), but it is marginally so in Maitzen's (\cite{M76})
system ($\Delta a=+0.008$). Catalano et al. (\cite{C98}) have found a marginal
infrared variability in the $J$ band with a period of 2.78 days, which would
favour the Ap classification. No clear-cut period can be obtained from our
data, at least under the assumption of an orbital motion. Would this star be a
{\it bona fide} magnetic Ap, the $RV$ variations could be attributed to spots
and rotation. However, many indications rather support the Am classification:
Babcock (\cite{B58}) did not find a significant magnetic field, Abt \& Morrell
(\cite{AM95}) classified this star as Am and Leroy (\cite{L95}) did not find any
significant polarization in it. Moreover, Ginestet et al. (\cite{G97}) also
classified it Am from near-IR spectra. The large $RV$ scatter remains unexplained;
perhaps an SB2 system seen nearly pole-on might produce such an effect. A
further in-depth investigation of this object would be necessary to clarify its
nature, but this is beyond the scope of this paper.

\begin{acknowledgements}
This work was supported in part by the Swiss National
Fondation for Scientific Research. The reduction of the data were made by the
late Dr. Antoine Duquennoy and by SU. We thank the numerous
observers who have contributed to this survey, especially Dr. J.-C. Mermilliod
and Mr. Bernard Pernier. We also thank Dr. No\"el Cramer, who had initiated
the CORAVEL measurements of some bright Ap stars.
We thank Dr. Thierry Forveille for having shared his ORBIT code for orbital
elements determination from visual and RV data.
This research has made use of
the SIMBAD database, operated at CDS, Strasbourg, France. It was
supported by the Swiss National Science Foundation.
\end{acknowledgements}

\end{document}